\documentclass[twocolumn]{aastex63}
\usepackage{graphicx}
\usepackage{subfigure}
\usepackage{amsmath}    
\usepackage{amssymb}    
\usepackage{bm}
\usepackage{url}
\usepackage{upgreek}
\usepackage{xspace}
\usepackage{longtable}
\usepackage{booktabs}

\usepackage{float}

\newcommand{\ppcc}{$\,$pc$\,$cm$^{-3}$\xspace} 
\newcommand{\tsys}{T$_{\rm sys}$\xspace}       

\newcommand{\bandiii}{{400 MHz}\xspace}
\newcommand{\bandiv}{{650 MHz}\xspace}
\newcommand{\bandv}{{1360 MHz}\xspace}
\newcommand{\bandl}{{1500 MHz}\xspace}
\newcommand{\bands}{{2000 MHz}\xspace}
\newcommand{\bandc}{{5400 MHz}\xspace}

\def\magn{XTE~J1810$-$197\xspace}

\def\tsys{$T_{\rm sys}$\xspace}

\def\jyms{Jy\hspace{0.05cm}ms\xspace}

\shorttitle{From Low-energy XTE~J1810$-$197 Bursts to FRBs}
\shortauthors{Lal et al.}

\begin{document}

\title{Low-energy Radio Bursts from Magnetar XTE~J1810$-$197: Implications for Fast Radio Bursts}

\correspondingauthor{Banshi Lal}
\email{blal@ncra.tifr.res.in}
\author[0009-0000-6605-3162]{Banshi Lal}
\affil{National Centre for Radio Astrophysics, Tata Institute of Fundamental Research, Post Bag 3, Ganeshkhind, Pune - 411007, India}
\correspondingauthor{Yogesh Maan}
\email{ymaan@ncra.tifr.res.in}
\author[0000-0002-0862-6062]{Yogesh Maan}
\affil{National Centre for Radio Astrophysics, Tata Institute of Fundamental Research, Post Bag 3, Ganeshkhind, Pune - 
411007, India}
\author[0000-0002-4441-7081]{Moaz Abdelmaguid}
\affil{Center for Astrophysics \& Space Science (CASS), NYU Abu Dhabi, PO Box 129188, Abu Dhabi, UAE}
\affil{Department of Physics, New York University, 726 Broadway, New York, NY 10003, USA}
\author[0000-0002-4629-314X]{Visweshwar Ram Marthi}
\affil{National Centre for Radio Astrophysics, Tata Institute of Fundamental Research, Post Bag 3, Ganeshkhind, Pune - 
411007, India}

\author[0000-0003-4679-1058]{Joseph D. Gelfand}
\affil{Center for Astrophysics \& Space Science (CASS), NYU Abu Dhabi, PO Box 129188, Abu Dhabi, UAE}
\affil{Center for Cosmology and Particle Physics (CCPP, Affiliate), New York University, 726 Broadway, New York, NY 10003, USA}

\author[0000-0003-4136-7848]{Samayra Straal}
\affil{Center for Astrophysics \& Space Science (CASS), NYU Abu Dhabi, PO Box 129188, Abu Dhabi, UAE}



\begin{abstract}

Magnetars are the leading candidate sources of fast radio bursts (FRBs). However, the observational probes of the connections between magnetars and FRBs are severely limited by the paucity of detection of highly energetic radio events from magnetars --- to date, only one radio burst as energetic as FRBs has been detected from a Galactic magnetar. Here, we present a detailed analysis of a large sample of low-energy bursts detected from the magnetar \magn, and probe their implications for FRB emission from magnetars. We report detection of over 97000 bright radio pulses from 242 observations of the magnetar \magn over 4.5 years and two decades in frequency (300 MHz to 6.15 GHz), using the Giant Meterwave Radio Telescope and the Green Bank Telescope, after its recent outburst onset in December 2018. We present detailed analysis of the burst fluence distributions and their trends with time as well as frequency, and the waiting time distribution. We show that \magn rapidly switches between pulsar-like and giant-pulse-like emission states, and magnetars like \magn remain viable and likely emitters of FRBs, in the form of giant-pulses with energies comparable to FRBs. We also demonstrate that the lack of the detection of an underlying periodicity in the bursts from repeating FRBs might be caused by emission across a wide range of spin phases.

\end{abstract}

\section{Introduction} \label{sec-intro}

Around $10 \%$ of the newborn neutron stars are expected to exhibit magnetic fields higher than $10^{14}$\,G \citep{popov_2010}. Neutron stars with such high magnetic fields are known as magnetars. Magnetars emit bright radiation across a range of wavelengths, mostly powered by the decay of their enormous magnetic fields \citep{Duncan_1992}. The majority of known magnetars are detected via their high energy emission, and have long spin periods, typically between 1.4 and 12\, seconds, but potentially extending to several tens of minutes or even hours. Historically, magnetars are comprised of two kinds of sources: Soft Gamma Repeaters (SGRs) and Anomalous X-ray Pulsars (AXPs). Furthermore, depending on the quiescent luminosity and dynamic range (DR: ratio of maximum and minimum observed X-ray luminosity), magnetars are categorized into two classes, persistent magnetars (quiescent luminosity $>10^{33}$ erg\,s$^{-1}$, DR$<$100) and transient magnetars \citep[quiescent luminosity $<10^{33}$ erg\,s$^{-1}$, DR$>$100;][]{Kaspi_2017}. Transient magnetars are detected during an outburst, like AXP XTE~J1810$-$197 \citep{Ibrahim_2004}, the first magnetar that was found to be emitting at radio frequencies. 

\par
\begin{figure*}[htp]
\centering
\label{fig:1}
\includegraphics[width=\textwidth]{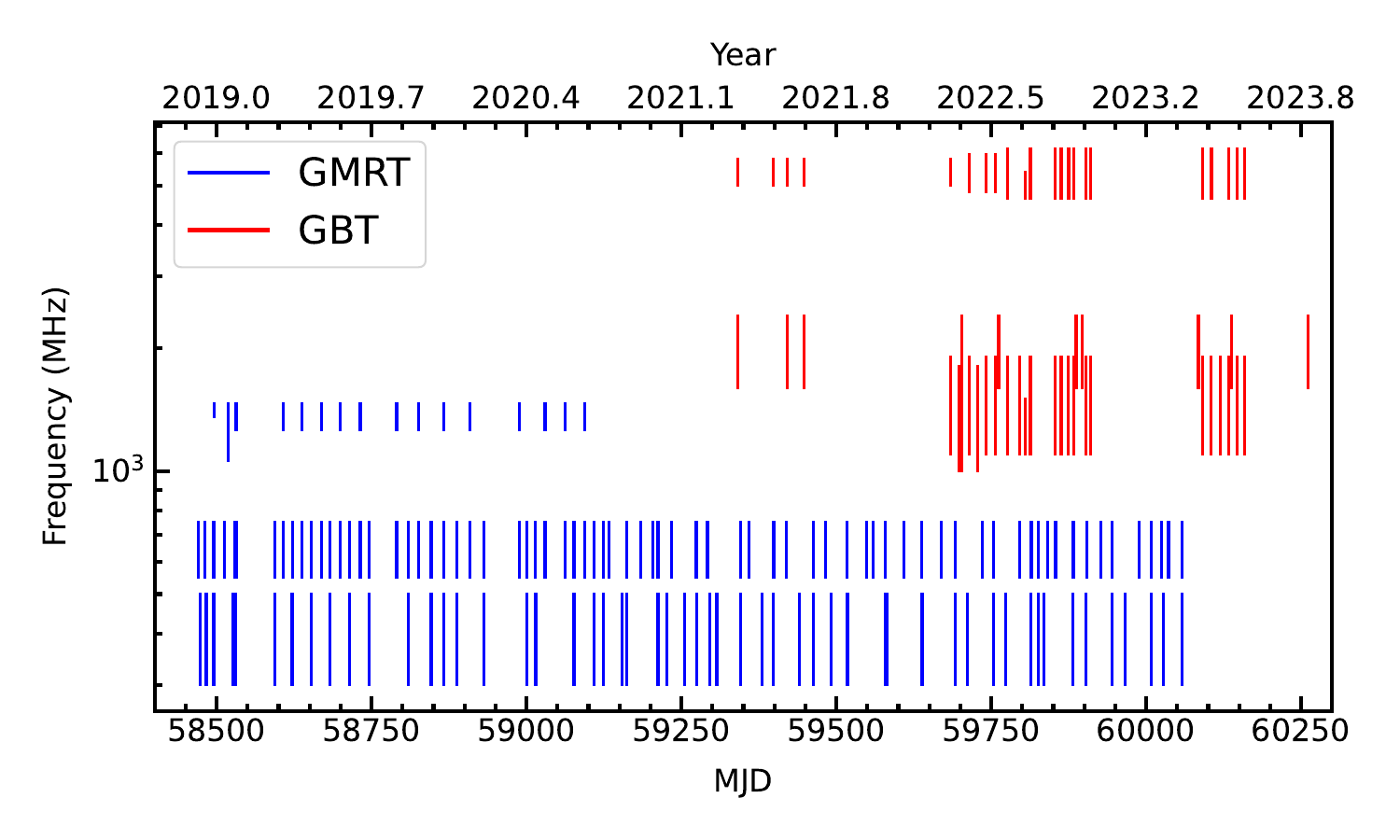}
\caption{The frequency coverage for the individual observations is shown with the observation date (MJDs as well as decimal years). The GMRT observations are shown in blue and the GBT ones are shown in red color.}
\end{figure*}

Magnetars are one of the leading candidate sources of fast radio bursts (FRBs) --- millisecond-duration bursts with very high energies. The dispersion measures (DM) of FRBs suggest their origin to be extra-galactic. Many theoretical models have been proposed to explain their progenitors and the emission mechanism \citep{Cordes_2019}. Magnetars have been discussed as FRB progenitors in several different contexts. For example, \cite{popov_2010} suggested that the Lorimer burst \citep{Lorimer_2007} can be a hyperflare from a magnetar. \cite{Lyubarsky_2014} discussed the interaction between a magnetar and an ambient wind nebula as an FRB emission mechanism. The only Galactic FRB, FRB 20200428D, has been observed from the magnetar SGR~1935+2154 \citep{Bochenek_2020, CHIME_2020}. FRB 20201020A shows pulse substructures \citep{Pastor_Marazuela_2023} similar to that seen in the bursts from magnetar \magn \citep{Maan_2019}. 

\par
\magn is a magnetar discovered in 2003 during an outburst \citep{Ibrahim_2004} in x-rays. This is the first known magnetar to show radio emission\citep{camilo_2006}. The radio flux density of XTE~J1810$-$197 decreased with time and it became abruptly undetectable in 2008 \citep{Camilo_2016}. A second outburst from this magnetar was detected in late 2018 \citep{lyne_2018, Maan_2019}. After this outburst, XTE~J1810$-$197 has been regularly monitored with different telescopes including the Giant Meterwave Radio Telescope (GMRT) and the Green Bank Telescope (GBT). Our monitoring observations using GMRT started in December 2018. Some of the results from this monitoring, e.g., time variability of the average profile, flux density, spectral index as well as the broadband spectral shape, were reported in \cite{Maan_2022}. In this work, we present a study of the bright single pulses with signal-to-noise ratio (S/N) $>$ 13 from the magnetar \magn, and probe their implications for the possible magnetar-FRB connection. So far, other than FRB 20200428D, no other FRB has been detected from the Galactic magnetars. Here, we have also probed if FRB-like emission could be expected from \magn, by studying the energetics and other properties of the radio bursts from this magnetar observed over a time span of around 4.5 years at various frequencies.
\par
The rest of the paper is arranged as following. Observation and data reduction methods, burst energetics, and waiting time distribution are described in sections ~\ref{sec-obs_data_reduction}, ~\ref{sec-burst_energy}, and ~\ref{sec-waiting_time}, respectively. We discuss our results and their implications in Section\ref{sec-discuss}, followed by conclusions in Section\ref{sec-conclusion}.

\section{Observations and Data Reduction} \label{sec-obs_data_reduction}
\setcounter{table}{0}
\begin{table*}[t]
    \centering
    \caption{Observation details}
    \label{tab:observation_details}  
    \begin{tabular}{c c c c c c c c}
        \hline
        \hline
       
        \textbf{Telescope} & \textbf{Frequency} & \textbf{Frequency} & \textbf{Central} & \textbf{Sampling} & \textbf{Number} & \textbf{Observation} & \textbf{Single Pulses} \\
        \textbf{} & \textbf{Band} & \textbf{Range}$^{*}$ & \textbf{Frequency}$^{*}$ &  \textbf{Time} & \textbf{of} & \textbf{Duration} & \textbf{above} \\
        \textbf{} & \textbf{} &\textbf{(MHz)} & \textbf{(MHz)} & \textbf{(\textmu s)} & \textbf{Observations} & \textbf{(Hours)} & \textbf{13 sigma} \\
        \hline
        GMRT & {3} & 300-500 &  \textbf{400} & 655.36 & 61 & 35.3 &  16311\\
        GMRT & {4} & 550-750 & \textbf{650} & 163.84 & 104 & 73.9 & 43139 \\
        GMRT & {5} & 1260-1460 & \textbf{1360} & 163.84 & 20 & 15.1 &  10987\\
        GBT & {L} & 1000-1900 & \textbf{1500} & 40.96 & 23 & 6.4 &  10127\\
        GBT & {S} & 1600-2400 & \textbf{2000} & 40.96 & 11 & 3.6 &  8281\\
        GBT & {C} & 4650-6150 & \textbf{5400} & 21.84 & 23 & 5.4 &  8422\\
        \hline
        \textbf{GMRT+GBT} & \textbf{} & \textbf{300-6150} &  & \textbf{} & \textbf{242} & \textbf{139.7} & \textbf{97411} \\
        \hline
    \label{tab:1}
    \end{tabular}
    \vspace{0.5em} 
    {\footnotesize $^{*}$ These are typical frequency ranges and central frequencies, and the actual numbers differ slightly for a few observations.}
\end{table*}

\subsection{Observations} \label{sec-obs}
We have used GMRT and GBT for our observations. GMRT is an array of 30 dishes, each with a diameter of 45 meters, spread over 25\,km near Pune, India. Especially after the recent upgrade \citep{YG_2017}, it is one of the largest and most sensitive telescopes at low radio frequencies. GMRT's incoherent and phased array beams provide high-sensitivity observations for compact objects, like pulsars,  magnetars and FRBs. Regular monitoring of magnetar \magn has been going on since December 2018. 
\par
GBT is the world's largest single-dish, fully steerable 100\,m diameter radio telescope in Green Bank, West Virginia, USA. The unblocked aperture and good surface accuracy of GBT provide high sensitivity for observations across a large frequency range of 0.3 to 116\,GHz.
\par
For this work, we have used 242 observations taken between December 2018 and March 2023 using GMRT (at \bandiii, \bandiv and \bandv) and GBT (at \bandl, \bands and \bandc) as mentioned in Table~\ref{tab:1}. A summary of the observations, in terms of frequency coverage with the observation date is shown in Figure~\ref{fig:1}. We have used observations that span a factor of 20 in frequencies (from 300 MHz to 6000 MHz) and cover more than 4.5 years in time.

\subsection{Data Reduction and Identifying Single Pulses}  \label{sec-data_red_single_pulses}

The typical data we get from radio telescopes are generally contaminated by radio frequency interference (RFI). Various measures have been taken to identify and excise RFI wherever possible. For GMRT data, {\fontfamily{qcr}\selectfont \texttt{RFIClean\footnote{\url{https://github.com/ymaan4/RFIClean}}}} \citep{Maan_2021_rficlean} is used. RFIClean also converts the native GMRT data format to SIGPROC filterbank format, which is then readable by packages such as the Pulsar search and analysis toolkit \citep[PRESTO;][]{Ransom_2001} and PSRCHIVE \citep{PSRCHIVE_2004}.

The raw GBT data were converted from PSRFITS to SIGPROC filterbank format using DSPSR's \texttt{digifil} routine. The output filterbank files were then sub-banded using SIGPROC's \texttt{dedisperse} routine, with 256 subbands and 32 bits per sample output. Then, \texttt{digifil} was again used to reduce the number of bits per sample to 8.

The above data reduction procedure results in SIGPROC filterbank format data for GMRT as well as GBT. For further RFI excision, {\fontfamily{qcr}\selectfont \texttt{rfifind}} from PRESTO is used, which identifies the RFI contaminated frequency channels and time sections and creates a mask file. In any further processing of the data, the mask file is used to exclude the RFI contaminated parts of the data. To obtain the folded, average profile of the magnetar, {\fontfamily{qcr}\selectfont \texttt{prepfold}} from PRESTO is used along with the above mask and ephemeris of \magn. The output of this command is written in a file with extension {\fontfamily{qcr}\selectfont \texttt{pfd}} which contains partially folded and frequency resolved average profiles as well as information like the best S/N and the corresponding DM and period. For an even better estimation of the period, DM, S/N, and the folded profile width, {\fontfamily{qcr}\selectfont \texttt{pdmp}} from PSRCHIVE is used on the {\fontfamily{qcr}\selectfont \texttt{pfd}} file.

To search for the bright single pulses, we first obtained dedispersed time series (DM=178.85\,\ppcc) using {\fontfamily{qcr}\selectfont \texttt{prepdata}} from PRESTO. We have used the DM of 178.85\,\ppcc obtained from a previous study of the magnetar utilizing some of the data used here \citep{Maan_2022} and also verified that this DM maximizes the S/N of our narrowest pulses. Moreover, the typical narrowest pulse widths of 2\,ms at 650\,MHz implies that a DM incorrect by as much as $\pm$0.3\,\ppcc will not smear the pulse across 200\,MHz bandwidth much to reduce its detection S/N significantly. At the higher frequency bands, the allowed uncertainty in the DM is much higher. We have separately assessed the best DMs obtained for the narrowest pulses across our observations, and found those to be consistent with our above choice of DM within the associated uncertainties. The single pulses with S/N greater than 8 and a maximum width of 50 ms are identified from the dedispersed time series using a slightly modified version of {\fontfamily{qcr}\selectfont \texttt{single\_pulse\_search.py}}. The {\fontfamily{qcr}\selectfont \texttt{single\_pulse\_search.py}} output gives different pulse parameters for the identified candidates: DM, pulse arrival time, S/N, and width. To estimate the best S/N and the corresponding width, {\fontfamily{qcr}\selectfont \texttt{single\_pulse\_search.py}} uses boxcar match-filtering, which has trial widths with non-uniform spacings (boxcar widths of 1, 2, 3, 4, 6, 9...300 samples). For the identified candidate pulses, we conduct a more refined parameter estimation using PSRCHIVE. We extract data around the pulse arrival time from the original Sigproc filterbank file, convert the extracted data to PSRFITS format using DSPSR, and then use \texttt{pdmp} from PSRCHIVE to estimate the best S/N and the corresponding pulse width and DM estimates. 

\begin{figure}[htp]
    \centering
    \hspace*{-0.6cm}
    \includegraphics[width=1.3\columnwidth,height=0.35\textheight]{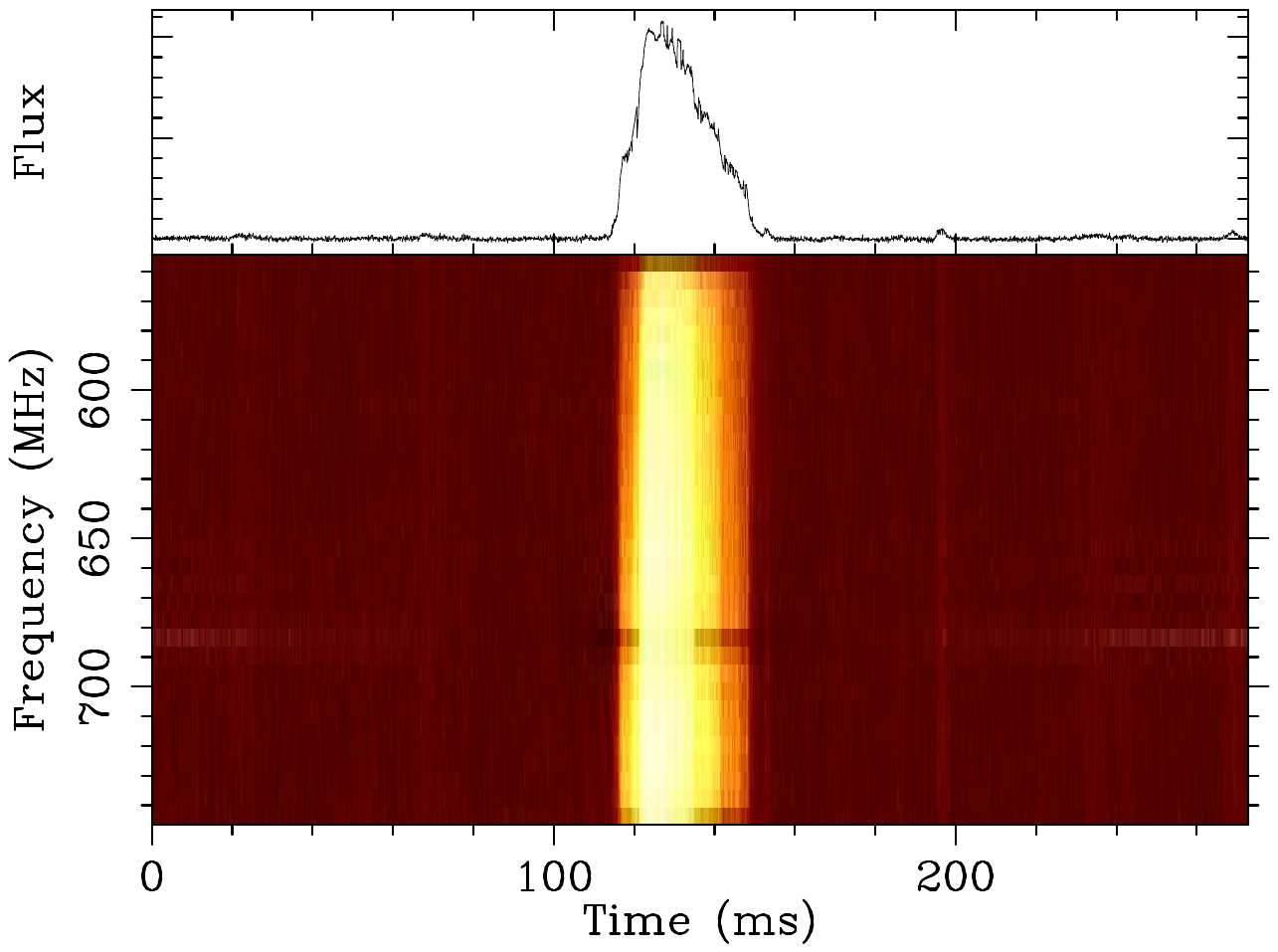}
    \caption{One of the brightest detected pulses: the bottom panel shows the dynamic spectrum de-dispersed at a DM of 178.85\,\ppcc, while the top panel shows the band-averaged pulse with flux density in arbitrary units.}
    \label{fig:2}
\end{figure}

\subsubsection{Candidate Filtering}

Although RFI mitigation was performed using {\fontfamily{qcr}\selectfont \texttt{RFIClean}} and {\fontfamily{qcr}\selectfont \texttt{rfifind}} which employ different methods to identify and mask or modify the contaminated parts of the data, some faint RFI might still escape identification. In addition to any such remnant RFI, other systematics, e.g., slow baseline variations, non-guassian noise characteristics, etc., could give rise to false positives while searching for single pulses. Thus, to identify genuine single pulses, additional filtering criteria was applied based on the best estimated DM. The \texttt{pdmp} output gives the best DM for every single pulse within the specified DM range which is between 168 and 188 \ppcc in this case. The best DM for genuine pulses is expected to align closely with the true DM (close to 178.85 \ppcc), whereas DM for RFI-originated candidates would be significantly away from true DM (ideally at 0 or very low DM). So we identified and removed such candidates with significant deviations from the true DM, and then left with only the genuine bright pulses.
\par
For the pulses with substructures, sometimes the \texttt{single\_pulse\_search.py} identifies different components as different pulses. In such cases, a single pulse might get counted more than once. We have examined such cases and if there is an overlap of more than 50 percent between two pulses (i.e., the pulse arrival time for i$^{th}$ pulse falls within the identified width of the (i+1)$^{th}$ pulse or vice-versa), then we choose only the pulse with higher S/N.

\par
After employing the above filtering criteria, the remaining pulses are used for further analysis. 

\begin{figure*}[htp]
    \centering
    \begin{minipage}[t]{0.49\linewidth}
        \centering
        \includegraphics[width=\linewidth]{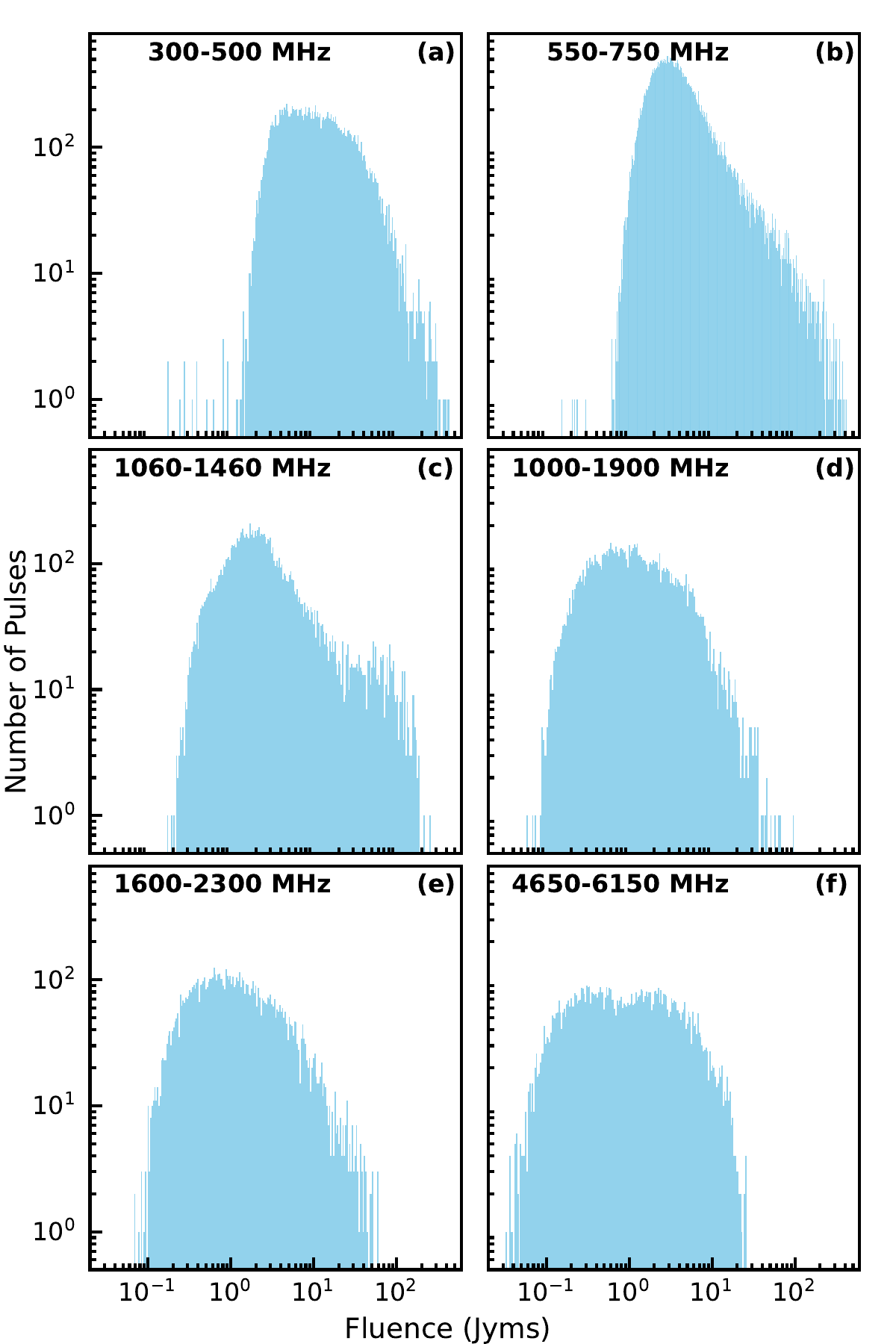}
    \end{minipage}
    \hfill
    \begin{minipage}[t]{0.49\linewidth}
        \centering
        \includegraphics[width=\linewidth]{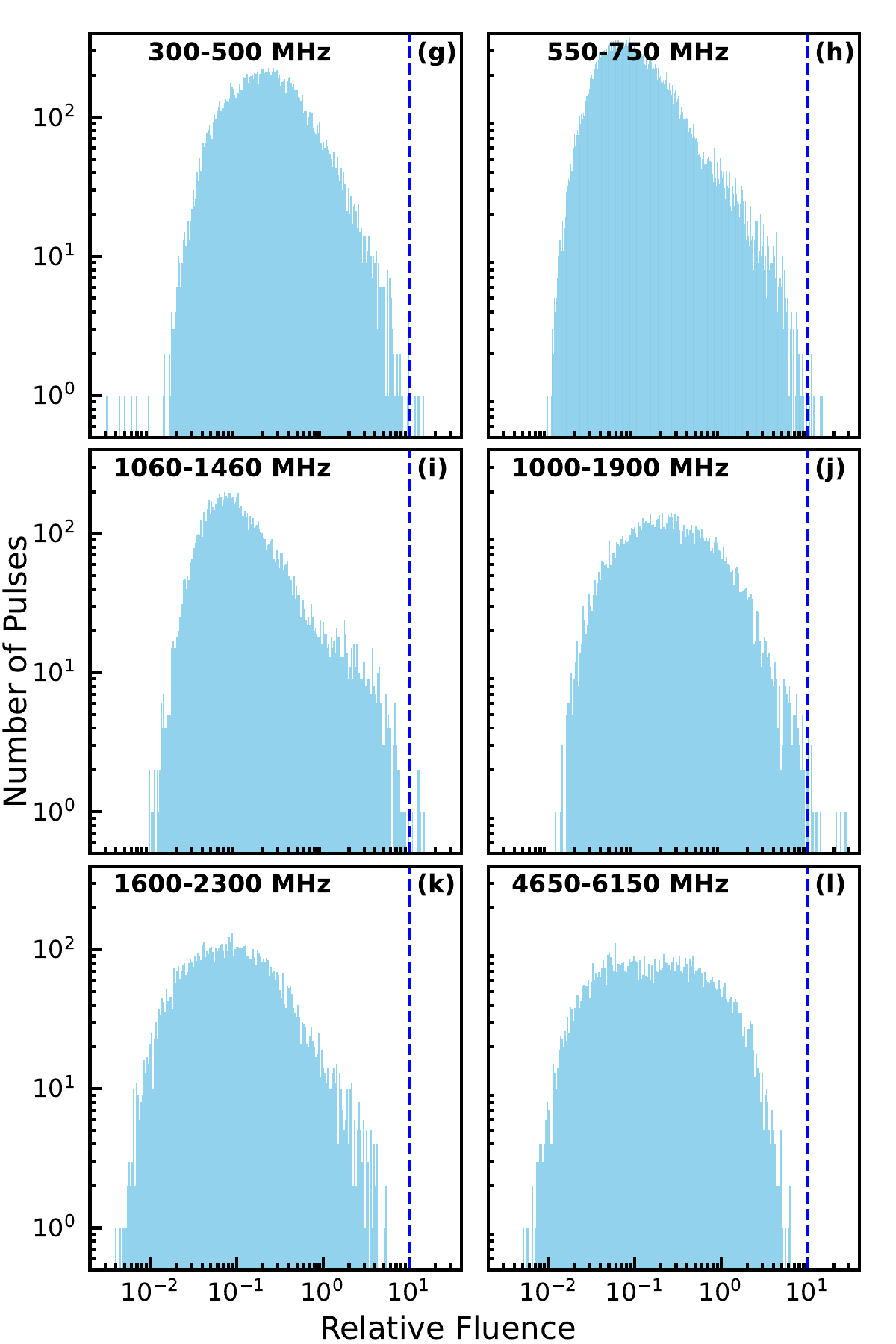}
    \end{minipage}
    \caption{Left set of panels: The fluence histograms using all the pulses detected at different frequencies are shown in different panels. The different subplots,  a, b, c, d, e and f correspond to \bandiii, \bandiv, \bandv, \bandl, \bands, and \bandc, respectively. Right set of panels: The relative fluence histograms at different frequencies. The blue vertical line, at a relative fluence of 10, is to distinguish normal pulses from giant-pulses. Giant pulses are defined as pulses with a relative fluence of equal to or greater than 10.}
    \label{fig:3}
\end{figure*}

\section{Burst Energetics} \label{sec-burst_energy}

\subsection{Flux density and fluence estimation} \label{sec-flux_density}

For GMRT observations, we have used the procedure described in \citet[section 3.1]{Maan_2022} to estimate the System Equivalent Flux Density (SEFD or \tsys/G). The estimated SEFD per GMRT dish as a function of frequency is fitted by a polynomial, and the fitted polynomial is then used to compute SEFD for the number of dishes used in the individual observations. For GBT, to estimate the SEFD, receiver temperature and gain are used as specified by the observatory\footnote{\url{https://www.gb.nrao.edu/~fghigo/gbtdoc/sens.html}} (For L, S, and C bands, \tsys = [20K, 22K, 25K], and Gain = [2, 1.9, 1.87]). The sky temperature towards the magnetar \magn is estimated at different frequencies using the package {\fontfamily{qcr}\selectfont \texttt{skytempy\footnote{\url{https://libraries.io/pypi/skytempy}}}}. The radiometer equation \citep{2004hpa..book.....L} is then used to compute the flux density and fluence of the individual single pulses. The radiometer equation uses the above estimated SEFD, the best estimates for S/N and pulse-width, and the observing bandwidth. We have used S/N and pulse width obtained from \texttt{pdmp} as the best estimates. 
\par
We have detected more than 97000 bright pulses ($S/N>13$), with a maximum fluence close to 0.5 K\jyms. We have chosen pulses with PRESTO S/N$>$13 for our analysis to ensure completeness of our single pulse sample\footnote{A detailed completeness analysis (Kumar et al. 2025; under preparation) suggests that a sample of single pulses with detection S/N$>$8 would be 90\% complete.}. One of the brightest pulses is shown in Figure~\ref{fig:2}. The number of pulses detected in the individual frequency bands are presented in Table 1. The fluence distribution of all the pulses at different frequencies are shown in the left panel of Figure~\ref{fig:3}.
\par
We also present the relative fluence, defined as the single pulse fluence divided by fluence of the folded profile from the respective observation. Note that the relative fluence is equivalent to the relative flux density, i.e., the ratio of period-averaged flux density of a single pulse to that of the folded profile. The distribution of the relative fluence for all the pulses from all observations at different frequency bands are shown in the right panels of Figure~\ref{fig:3}.

\par

\subsection{Fluence distribution characterization and evolution} \label{sec-distributions}

\begin{figure*}
    \centering
    \includegraphics[scale=0.38]{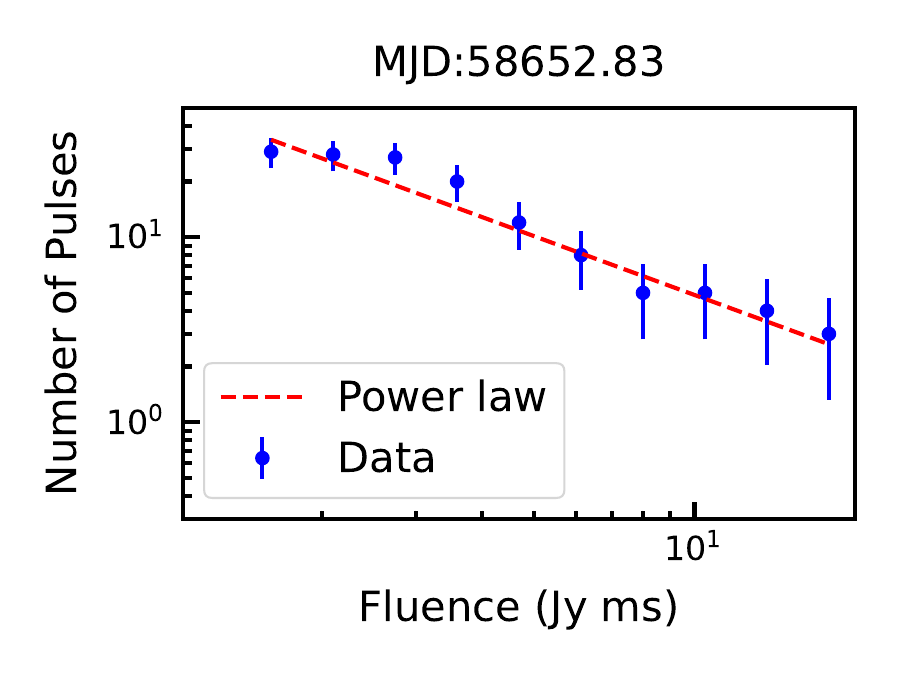}
    \hfill
    \includegraphics[scale=0.38]{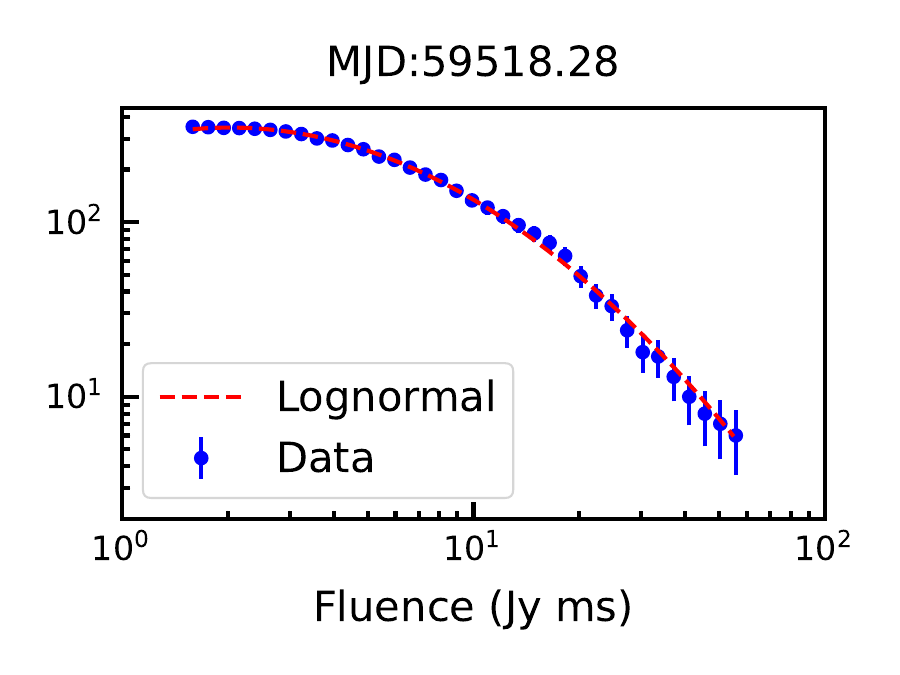}
    \hfill
    \includegraphics[scale=0.38]{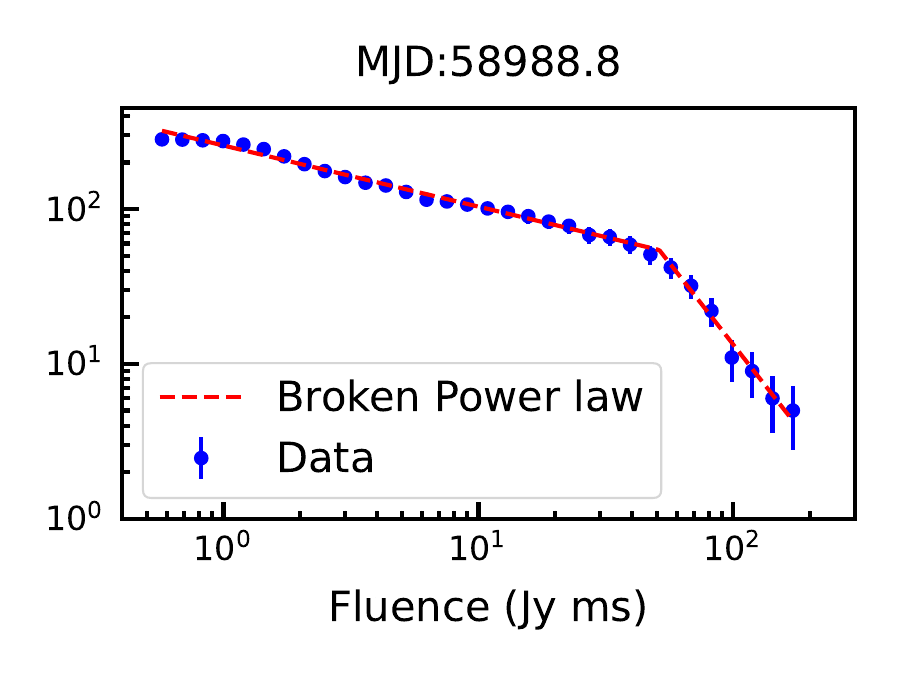}
    \includegraphics[scale=0.4]{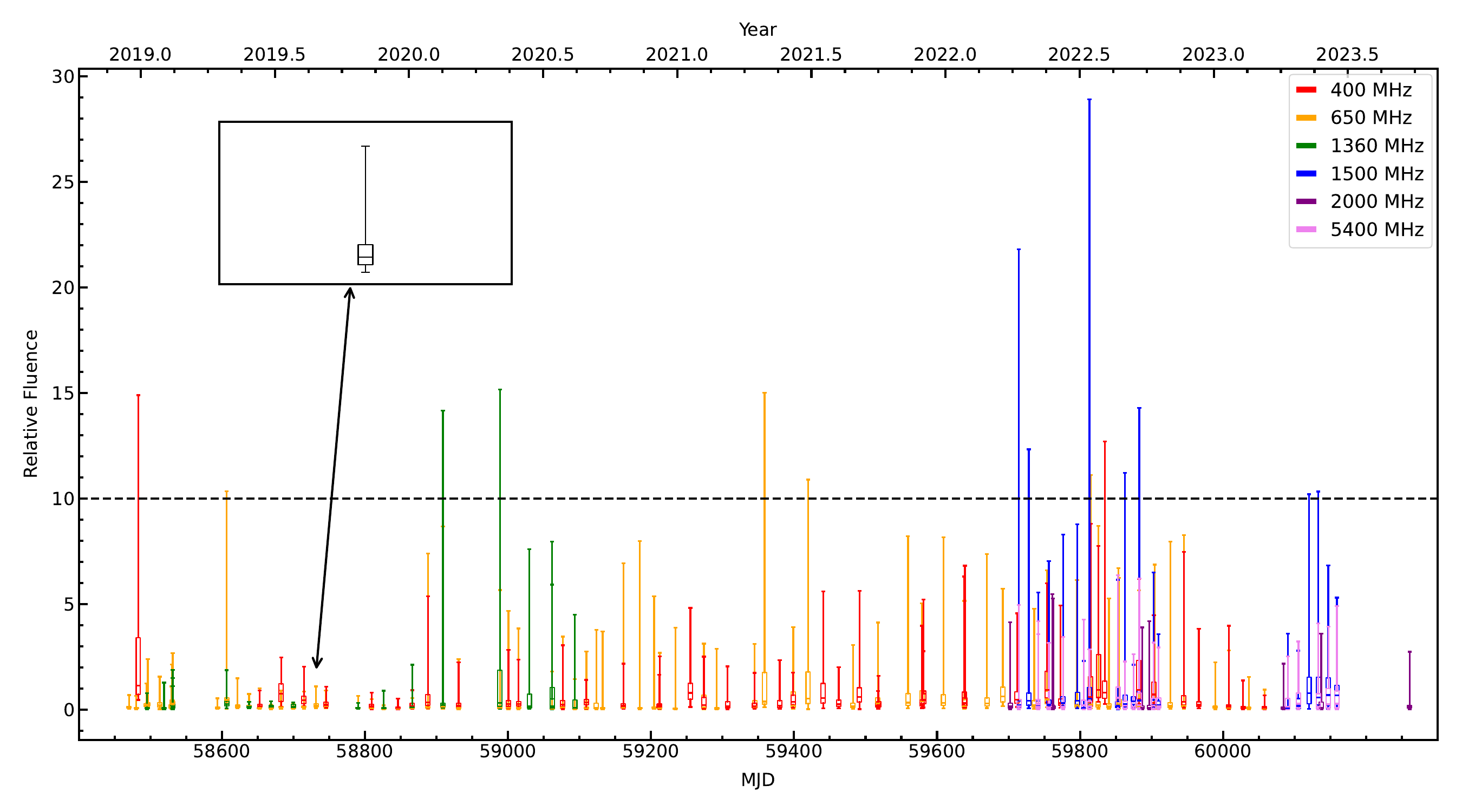}
    \caption{The bottom panel shows the relative fluence distributions for all the observations using box plots. The zoomed-in figure illustrates the box plot for a typical observation, where the top and bottom bars on the box plot are maximum and minimum relative fluences, the top and bottom of the box itself are the 75 and 25 percentile, and the line inside the box is the median of the relative fluence distribution. The black dashed line at 10 is for giant pulses. A log scale version of this figure is shown in Appendix~\ref{box_plot}. Top: A few examples of cumulative fluence distributions at different epochs is shown. The blue points represent the observed number of pulses with Poissonian uncertainty and the red dashed line represents the fitted model. At the top of these plots, the epoch of observation is mentioned.}
    \label{fig:4}
\end{figure*}

The relative fluence distributions for individual observations are shown in the bottom panel of Figure~\ref{fig:4} using box plot. It is apparent that the maximum relative fluence changes rapidly with time. The distributions also show some variation with frequencies as well. We have noticed that not only the range of relative fluence observed within the individual observations but the fluence distribution shape also changes from epoch to epoch. The top panel of Figure~\ref{fig:4} shows examples of different cumulative fluence distributions observed at three different epochs.

\par
Earlier studies suggest that the fluence distributions of FRBs \citep[e.g.,][]{zang_2019,Wu_2025} and giant pulses from pulsars \citep[e.g.,][]{bera_2019} typically follow a power-law or power-law with a break. On the other hand, the regular and occasionally sporadic bright pulses from pulsars typically follow a lognormal distribution \citep[e.g.,][]{Burke-Spolaor_2012}. In some cases, where some of the regular single pulses as well as giant-pulses from a pulsar are detectable, the low energy part of the distribution could follow a lognormal distribution and the tail-end might follow a power law, and hence, such distributions would be better characterized by a modified lognormal power-law (MLP) model. The MLP model includes a power-law tail to the lognormal distribution and we use the formulation for such a model as described by \citet{Basu_2015}. So, to cover all of the above possibilities, we characterize the fluence distributions from individual observations using four models: (i) power-law, (ii) lognormal, (iii) MLP, and (iv) a broken power-law, and then choose the model that fits the data best.
\par
We have fitted these four models to each of the distributions using Scipy's {\fontfamily{qcr}\selectfont \texttt{curve-fit}}. Then we choose the best model using Akaike’s information corrected criterion (AICC). AICC considers the goodness-of-fit (chi-squared), the complexity of the model (number of fitted parameters), and the number of data points to be fitted. The AICC values and reduced chi-squared values for each of the models are listed in Appendix~\ref{sec-appendix_table}. We note that at some epochs, a few models result in reduced chi-squared to be less than 1 (see Appendix~\ref{sec-appendix_table}), indicating over-fitting. However, AICC still helps in choosing the best model depending on the criteria mentioned above. To estimate the uncertainty in the binned number of pulses in the distributions, we have assumed Poissonian statistics. The Poisson statistic predicts square-root uncertainty for large numbers and asymmetric uncertainty for small numbers. However, {\fontfamily{qcr}\selectfont \texttt{curve-fit}} does not allow usage of the asymmetric uncertainty while fitting. So, our choice of the model that fits the data best is based on fitting that considers only symmetric Poissonian uncertainties. After selecting the best model, the model parameters are characterized using Markov chain Monte Carlo (MCMC). The MCMC approach incorporates asymmetric uncertainties for pulse counts below 30 and the square root of the counts for larger numbers.

\begin{figure*}[htp]
\centering
\hspace*{-0.8cm}
\includegraphics[scale=0.55]{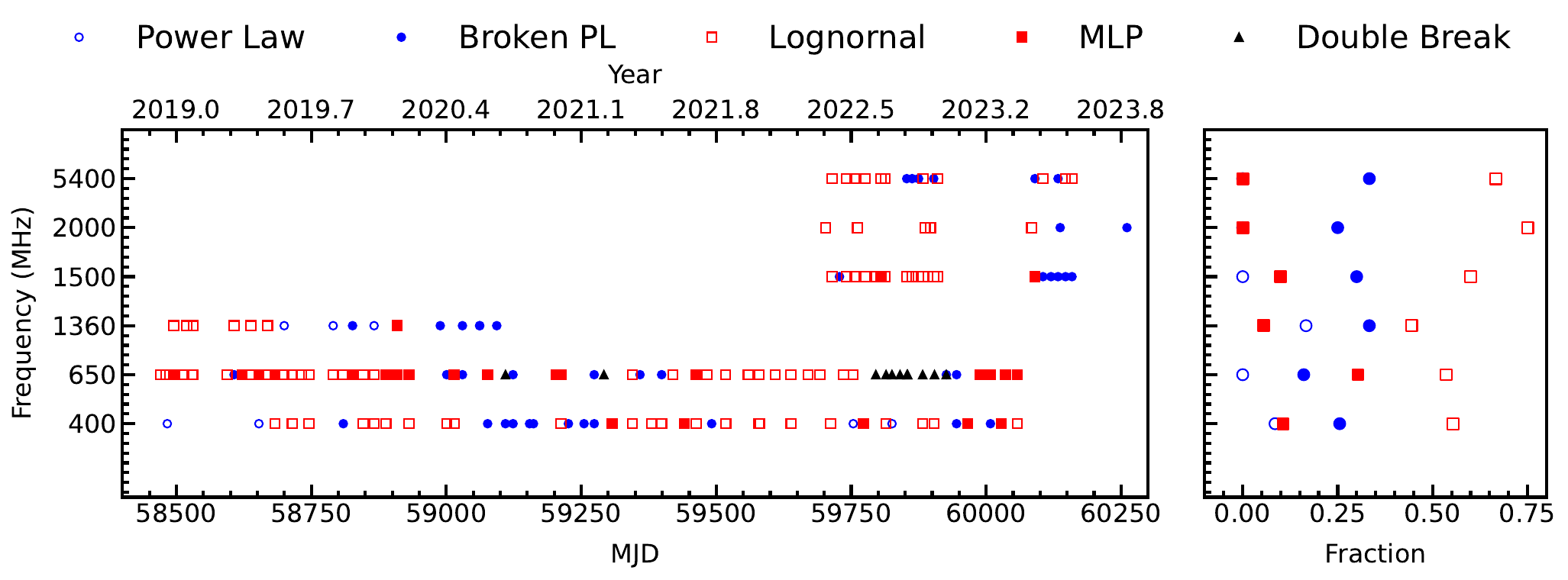}
\caption{Left: The best-fitted distribution is shown as a function of Epoch (in units of decimal year on the top and MJD in the bottom horizontal axis). The different symbols represent different models, as shown in the legends at the top. Right: The Fractions of the number of observations fitted with different models are shown, with the same color-symbol mapping for different models as in the left panel.}
\label{fig:5}
\end{figure*}
\par

\begin{figure*}[htp]
\centering
\includegraphics[width=\linewidth]{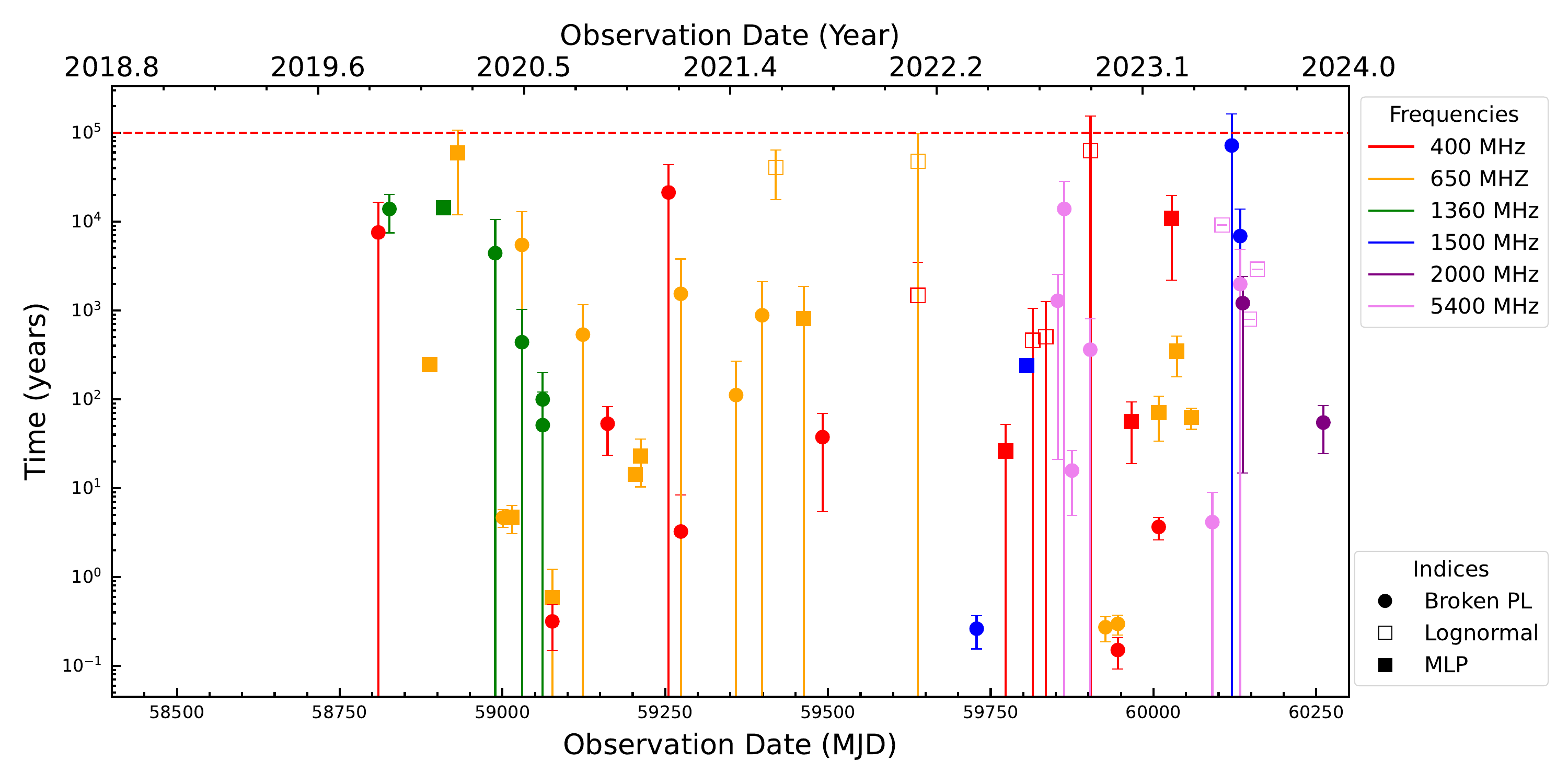}
\caption{The predicted timescale for emission of a burst with fluence 1.5\,M\jyms is plotted as a function of observation date, for all the epochs for which the timescale $<10^5$ years. The different colors represent the different frequency bands, and the different marker-shapes (symbols) are for the different best-fitted models, as indicated by the legends on the right side.}
\label{fig:6}
\end{figure*}

\par
For the MCMC parameter estimation, we have used the \texttt{emcee} package \citep{emcee} with the best fit estimates from the \texttt{curve\_fit} as the initial guesses for the chosen best model. The asymmetric uncertainties on the data points are used while defining the likelihood functions. The MCMC estimation gives a posterior distribution for the model parameters (a few cases are shown in Appendix~\ref{sec-appendix_C}), and the respective medians are chosen as the best-estimated parameters along with the associated errorbars deduced from the 84 and 16 percentiles. The posterior distributions for the model parameters and any covariances between them were manually examined for each of the observations, and fitted parameters were found to be well constrained for all the epochs. A summary of the best-fitted model with observation date is shown in Figure~\ref{fig:5}. All of the distributions were fitted very well with one of these four models. However, for a few cases, the fit was found to deviate in the last few bins at the tail end (due to a sudden decrease in the number of pulses). Such deviations do not significantly impact the goodness-of-fit, however, it is important to take such cases in account properly for the further analysis below. Thus, for these few cases, we have fitted a power-law with two breaks, which is described in Appendix~\ref{sec-appendix_A}.
It is noticed that the observations fitted with a power-law with double break are clustered around in a narrow range of epochs and primarily at one of the frequency bands (Figure~\ref{fig:5}). It might suggest that the magnetar was in a particular emission state during that time.
\par

\subsection{Timescale for FRB-like emission} \label{sec-timescale}

As described in Section\ref{sec-distributions}, the fitting procedure characterizes the detected number of pulses as a function of fluence in the cumulative distributions. Taking into account the observation duration, the cumulative distributions as well as the corresponding fits can be changed to burst-rate as a function of fluence. To compute the burst-rate at a desired very high fluence, e.g., a fluence where the energy would be comparable to that of FRBs, we can extrapolate the fitted distribution. For a particular observation, the extrapolated burst-rate provides the timescale needed to observe at least one pulse of a desired fluence, assuming the magnetar remains in that particular emission state.

To estimate the timescale as well as its associated uncertainty, we utilize the model parameter distributions resulted from MCMC sampling. By using these distributions, we compute the distribution of timescales for the individual observations. The robust mean and standard deviation (estimated by iteratively excluding outliers beyond the 4-sigma limit, using \texttt{sigma\_clipped\_stats} from \texttt{scipy}) of these timescale distributions are then used as the required timescale with the uncertainty.
\par
We have computed the timescale corresponding to a fluence of 1.5\,M\jyms for all the observations. The only known Galactic FRB, FRB 20200428D was detected with a fluence of 1.5 M\jyms at 1.4 GHz, and its energy is comparable to that of faint FRBs. Figure~\ref{fig:6} shows the timescale estimates for all the observations at different frequencies with the observation date. Only the timescale estimates that are smaller than $10^5$\,years, i.e., the typical active lifetime of magnetars, are shown. The presence of timescale estimates shorter than $10^5$ years at multiple epochs suggests that \magn could potentially produce an FRB-like burst during its active lifetime.

\par
We have noticed that the epochs with distributions fitted with only a power-law as the best model often had a low number of total pulses (less than 70), so we have excluded all those cases from Figure~\ref{fig:6}. For the timescale estimates of less than 100 years, we further scrutinized the tails of the distributions for individual epochs (see Appendix~\ref{sec-appendix_B}). The separate modeling of the distribution-tails make the timescale estimates longer at some of the epochs and shorter at some other epochs, however, the overall minimum timescales remain of the same order as those in Figure~\ref{fig:6}.

\section{Burst waiting time and duration} \label{sec-waiting_time}

\begin{figure}[htb]
    \centering
    \hspace*{-0.8cm}
    \includegraphics[scale=0.6]{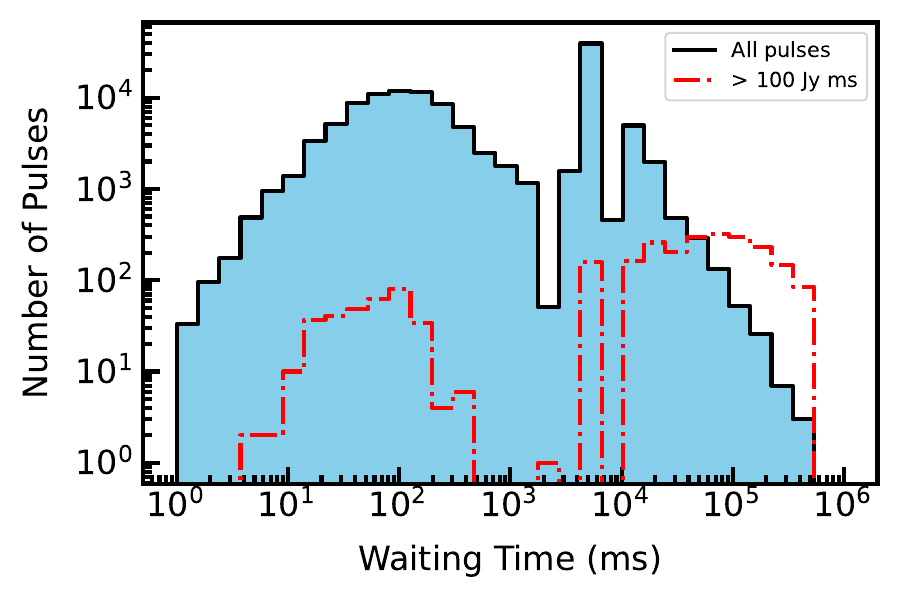}
    \caption{The waiting-time distribution for the bright pulses from magnetar \magn. The blue distribution represents the distribution using all the pulses, while the red colored one is the waiting-time distribution only for the pulses which have fluence $>$ 100 \jyms.}
    \label{fig:7}
\end{figure}

Our large number of bright pulses allows us to study the waiting time distribution. The waiting time is defined as the difference between the arrival times of consecutive pulses within an observing session. The cumulative waiting time distribution for all the bright pulses across all frequencies and observing sessions is shown in Figure~\ref{fig:7}. 

The distribution shows many peaks, of which the first one is around 100 ms and the subsequent peaks are at the magnetar's spin period of 5.54s and its harmonics. Here logarithmic binning has been used, which causes the peaks at the magnetar's spin period and its harmonics to merge together at adequately large bin sizes. We have also plotted the waiting time distribution of all the pulses with fluences greater than 100\,\jyms in Figure~\ref{fig:7}. The waiting time distribution for these brighter pulses shows two completely separate clusters. However, the overall charateristics are the same --- the first peak is around 100\,ms and the subsequent peaks are at the magnetar's spin period and its harmonics merged together due to large bin widths.

\begin{figure}[htp]
    \centering
    \hspace*{-0.5cm}
        \includegraphics[scale=0.48]{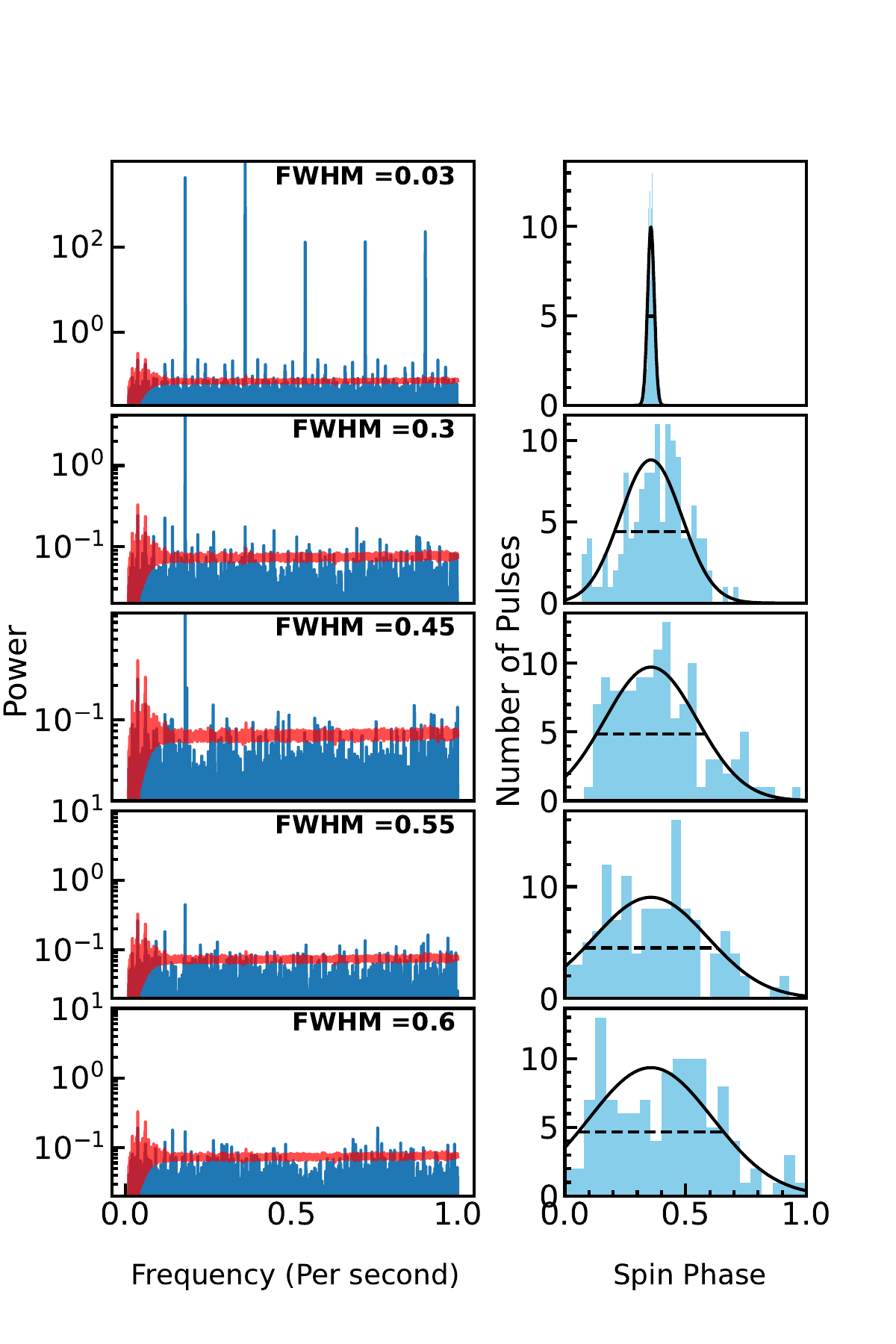}
    \caption{The panels on the left show the Lomb-Scargle periodograms for the pulses with fluence $>100$ \jyms, detected on 1st October 2020 at \bandiv. The right panels show the distributions of the pulse arrival times in the magnetar's spin-phase, corresponding to the periodograms shown on the left. The panels in the first row correspond to the observed arrival times of the bursts, while the distribution in the spin-phases are artificially spread in other rows. The red shaded lines in the left panels represent the 3.5-sigma significance levels, and the black dashed line in the right panels represents the FWHM of the spin-phase distribution.}
   \label{fig:8}
\end{figure}

\subsection{Periodicity Search} \label{sec-periodicity}
To understand the potential reasons for the lack of detectable periodicity in repeat bursts from FRBs, we use one of our monitoring observations, conducted on 1st October 2020 (MJD 59123). In this observation, we detected 130 pulses with fluence greater than 100 \jyms. The arrival times of these pulses are constrained well within the average profile window. The spin-phase distribution for these pulses, computed using the period of the magnetar \magn, shows a narrow distribution with a full-width half maxima (FWHM) of 0.03 (top right of Figure~\ref{fig:8}). The average pulse profile was also narrow in this observation, indicating that these pulses are spread over the full emission window of the magnetar \magn. The underlying periodicity of these bright pulses is easily detectable using the {\fontfamily{qcr}\selectfont \texttt{Lomb-Scargle (LS) Periodogram\footnote{\url{https://docs.astropy.org/en/stable/timeseries/lombscargle.html}}}} \citep{Lomb_1976, Scargle_1982, Lomb_scargle} on their arrival times. In the periodogram, the LS-power peaks at the magnetar's spin period and its harmonics, as shown in the top left panel of Figure~\ref{fig:8}. The significance of various peaks is estimated using a bootstrap method. Keeping the total number of pulses and the end-to-end time span the same as for the original pulses, 1000 random arrival times and the corresponding LS periodograms were generated. The distributions of LS powers at each of the frequency bins obtained from these 1000 periodograms were then used to estimate the mean and standard deviation, and the 3.5$\sigma$ significance level thus estimated is shown as the red shaded lines in the periodograms in Figure~\ref{fig:8} as a function of Frequency.

\par
To understand the effect of distribution of the arrival times across the spin-phase on detectability of the underlying periodicity, we artificially distributed the arrival times of the detected pulses from the above particular observing sessions in Gaussian windows of pre-specified widths. While redistributing the arrival times to effectively spread the spin-phases randomly across the Gaussian windows, the new arrival times of the individual pulses are kept closest to the original times. The LS periodograms of a few such realizations of the above procedure, along with the FWHM of the Gaussian window as well as the histograms of the redistributed spin-phases are shown in Figure~\ref{fig:8}. Note that the underlying periodicity becomes undetectable once the pulse arrival times are distributed in spin-phase windows of FWHM 0.55$-$0.6.

\section{Discussion} \label{sec-discuss}
We have presented the study of bright pulses from the magnetar \magn observed over a time span of $\sim$4.5\,years, and a frequency range spanning more than two decades (300-6150 MHz). Our focus has been primarily on the analysis of the energetics and other properties of the bright bursts, and their evolution with time and frequency. Below we discuss the implications of our results in terms of deciphering the nature of the bright pulses and probing the observational connections between magnetars and FRBs.
\subsection{Nature of the bright pulses}
Magnetar \magn emits very bright single pulses, sometimes with peak flux densities as high as a few hundreds of Jy. Similarly bright pulses known to be emitted from a handful of pulsars are called giant pulses --- conventionally defined as pulses having the period-averaged flux densities 10 times more than that of the average profile. Figure~\ref{fig:4} clearly shows that the magnetar occasionally emits giant pulses according to the above definition, especially at lower frequencies. 
\par
The combined fluence distributions from all pulses change with frequency, as shown in Figure~\ref{fig:3}. The magnetar emits more energetic pulses at lower frequencies than at higher frequencies. This might indicate that, statistically, the bursts from the magnetar have steep spectra, however, we note that the magnetar's spectral shape evolves with time \citep{Maan_2022} and our observations do not have a uniform frequency coverage throughout the monitoring campaign. 
\par
The distribution types themselves appear to change with time and frequency. When the magnetar is in a relatively quiet state (i.e., the observed maximum relative fluence is only a few), the fluence distribution is mostly of a lognormal type. At other times, the distributions require a power-law tail to be fitted (with the broken power-law and MLP models). Earlier studies show that the single pulses from normal pulsars typically follow lognormal distributions \citep[see, e.g.,][]{Mickaliger_2018,Burke-Spolaor_2012}, while the giant pulses follow a power-law distribution \citep[e.g.,][]{crab_2010}. So, the magnetar is potentially switching between pulsar-like and giant-pulse-like emission states and sometimes having a mixture of the two states. As the lognormal distribution has been found to be most commonly occurring at all the frequency bands (see the right panel of Figure~\ref{fig:5}), the magnetar appears to spend more time in a normal pulsar-like emission state than in a state where giant-pulses are emitted.
\par
Some distributions follow a power law with double breaks, which might raise a concern whether there is a break in other distributions too but outside the data span. However, we notice that majority of the distributions fitted with double breaks are clustered in time and frequency, as shown in Figure~\ref{fig:5}. This clustering might indicate a different emission state of the magnetar \magn during these epochs.
\par
Earlier studies have shown that the magnetar's radio spectrum changes with time, even with an evolving turnover \citep{Maan_2022}. A slow trend with frequencies is also apparent in Figure~\ref{fig:5}. For example, the frequency of the observations exhibiting lognormal distributions appears to decrease with time (from \bandv to \bandiv to \bandiii, between early 2019 and mid-2020). This evolution with frequency might also have contributions due to the change in the spectral behavior.

\subsection{Implications for FRB-like energetic emission from magnetars}\label{sec-frb-like energetics}
Magnetars are the leading candidate sources of FRBs. The detection of Galactic FRB from the magnetar SGR~1935+2154 confirms that some of the FRBs could be produced by magnetars. However, no further burst as energetic as FRBs has been observed from any of the Galactic magnetars so far. Nevertheless, some of the magnetars, like \magn, are known to be active in radio for extended durations and the low-energy bursts observed from them could educate us on what timescales emission of FRB-like energetic bursts could be expected from such magnetars, if at all. 
\par
Detailed modeling of the distributions of the detected bursts from individual epochs allows us to characterize the burst-rate as a function of fluence. We extrapolate the modeled burst-rate to a nominal fluence of 1.5\,MJy\,ms (same as that of the Galactic FRB at 1.4 GHz) and estimate the corresponding timescales. These estimates for the individual sessions for which the timescales are shorter than 10$^5$\,years are shown in Figure~\ref{fig:6}. Even after taking into account the caveats discussed in Section~\ref{sec-timescale} (e.g., exclusion of some epochs with small number of bursts and the need for modeling with double breaks at a few epochs), it is evident from Figures~\ref{fig:6} and \ref{fig:11} that, at several occasions, the magnetar is in an emission state where it could emit FRB-like energetic bursts at timescales even shorter than a year.
\par
There is an apparent trend in the expected timescale as well as the emission state of the magnetar. Until mid-2019, the emission state is such that the fluence distributions mostly follow a lognormal distribution, and there are no corresponding points on the timescale plot (i.e., Figures~\ref{fig:6}) as the expected timescale is larger than $10^5$ years.  Then, from late 2019 to the start of 2021, the fluence distributions primarily follow a power law distribution at the tail-end (either MLP or broken power-law), and the expected timescale is shorter than $10^5$ years. The timescale also decreases from late 2019 to late 2020, and then it increases (although not very systematically) again. If the magnetar had continued to evolve in the manner it does between late-2019 and late-2020, the timescale could have been even shorter, or it could already have emitted FRB-like energetic bursts. Then, from the start of 2021, the emission state is such that the fluence distributions mostly follow a lognormal distribution, and so on. In addition to the above potential slow trends, there are also huge, apparently uncorrelated variations in the expected timescales from one epoch to the other. Such variations could also result in the magnetar to occasionally emit a FRB-like energetic burst at a much shorter timescale in a favorable emission state.


\par
It is also evident from Figure~\ref{fig:6} that all the fluence distributions predicting timescales less than a few hundred years exclusively have a power-law tail (the broken power-law and MLP models). Assuming that the power-law tail is contributed by the giant-pulses implies that the magnetar is likely to emit FRB-like energetic bursts in the form of its giant pulses. 
\par
Giant-pulses from pulsars and the bursts at the brighter fluence end from repeating FRBs are known to follow a power-law distribution \citep[e.g., a power-law index of \textminus 1.6 to \textminus 1.8 is measured for FRB 121102][]{zang_2019}. The giant-pulses from the Crab pulsar exhibit a power-law energy distribution with an index $\approx -3$ \citep{bera_2019}. The timescale for a 1.5\,MJy\,ms fluence burst, estimated from the above distribution of Crab's giant-pulses, is a few hundred years! Unless rapid changes in the pulse-emission rate is found, the giant-pulse emission from Crab-like pulsars could explain only a tiny fraction of the observable FRBs. On the other hand, the magnetar \magn's is seen to change its emission state rapidly, with the power-law index of the distribution tails spanning a large range (\textminus 3.5 to \textminus 1.0) that covers the observed values for FRBs as well as pulsars. The ability to emit giant-pulses and such rapid changes in the emission states make magnetars much more favorable candidate sources of FRBs than Crab-like pulsars.
\subsection{FRB rate from the Galactic magnetar population} 

MLP and the broken power law provide reasonably short timescales for the fluence of 1.5\,MJy\,ms, as discussed in Section~\ref{sec-frb-like energetics}. The MCMC fitting provides parameters distribution, which is used to compute burst-rate distribution corresponding to fluence 1.5 \,MJy \,ms, for each observation. Using all the cumulative burst-rate distributions, we have determined the average cumulative burst-rate distribution for the observations best fitted with either MLP or broken power Law. The cumulative burst-rate distribution is then used to compute the cumulative timescale distribution for a fluence of 1.5\,MJy\,ms. The distribution gives a median expected timescale of $940$ years for \magn.

\begin{figure}[htp]
    \includegraphics[scale=0.5]{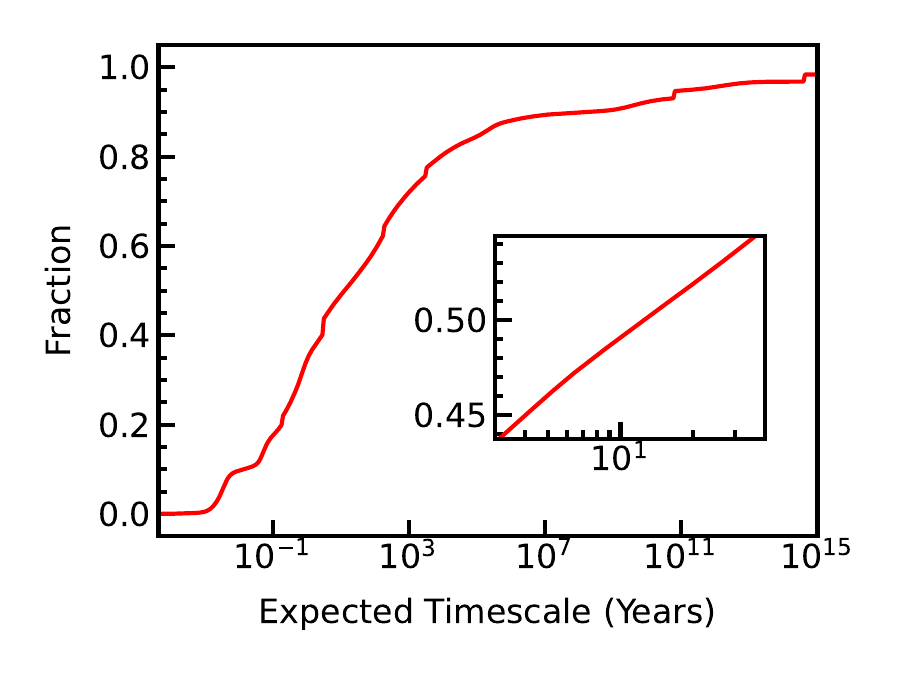}
    \caption{Cumulative expected-timescale distribution (for a burst with fluence 1.5\,M\jyms), obtained from the cumulative burst-rate distributions averaged for all the epochs best fitted using the MLP and broken power law models for \magn. The timescale distribution is further scaled appropriately (see text) to provide the estimate for the Galactic population of active magnetars. The inset shows a zoomed-in view around the median value.}
    \label{fig:cumu_timescale}
\end{figure}

According to a study by \cite{Gill_2007}, the magnetar birth rate is estimated to be 0.22 per century. If we consider the active lifetime of a magnetar to be $10^5$ years, the number of active magnetars in our Galaxy would be $\sim220$. Our analysis indicates that the magnetar XTE~J1810$-$197 follows MLP or broken power law in 34$\%$ of the observations. After considering the fraction of observations following MLP or broken power law and the above number of active Galactic magnetars, the median (i.e., at 50 percentile) of the scaled distribution, shown in Figure~\ref{fig:cumu_timescale}, suggests the expected timescale to detect a burst with a fluence of 1.5\,MJy\,ms from the Galactic magnetar population to be around $13$ years. We also note that the large timescales ($>10^9$\,years) in the cumulative distribution shown in Figure~\ref{fig:cumu_timescale} are contributed by a few distributions with very steep tail-ends.
\subsection{Waiting-time distribution and the underlying periodicity in FRBs}
For repeating FRBs, the burst-rate and waiting-time distribution are two of the well-studied properties. Repeating FRBs exhibit varying burst rates, often with periods of hyperactivity. The waiting-time distributions for many repeating FRBs show a bimodal nature, with the two peaks around 10-100 ms and 10-100 seconds e.g., for FRB 20201124A, the peaks are around 50 ms and 10 seconds \citep{Zhang_2022}, also for FRB 121102 \citep{Li_2021} at 3.4 ms, 70 seconds. \cite{Hu_2023} have studied 16 repeating FRBs and found bimodality in waiting time distribution for many repeating FRBs.
\par
The waiting-time distribution of the bright pulses from \magn also shows a bimodal nature, with peaks at around a hundred ms and a few tens of seconds. The bimodality becomes even clearer if we only use the higher fluence end of the burst population (see Figure~\ref{fig:7}). So, the waiting-time distribution of the low-energy bursts from \magn shows striking similarities with that of the repeating FRBs.
\par
The second peak in the waiting-time distribution of the magnetar bursts is caused by the magnetar's spin period and its harmonics. A search for periodicity in these bright pulses using {\fontfamily{qcr}\selectfont \texttt{Lomb-Scargle Periodograms}}, as shown in the top panels of Figure~\ref{fig:8}, easily recovers the magnetar's spin period. However, no such periodicity for FRBs has been found despite of several similar searches using statistically significant number of bursts. For example, \citet{Li_2021} made an unsuccessful attempt to find periodicity in bursts from FRB 20121102A in the range 1\,ms to 1000\,s. In Section\ref{sec-waiting_time}, we demonstrated that the underlying periodicity could become undetectable if the bursts are emitted in a wide range of spin phases. Using a particular observation with 130 bursts detected with fluence more than 100\,\jyms, we show that the underlying magnetar's spin periodicity would be undetectable even if the bursts are emitted in a range of 0.55$-$0.6 of the spin period. 
Thus, the bursts from repeating FRBs might be getting emitted at a wide range of spin-phases and thus limiting the detectability of the underlying periodicity. 
\par
Emission over as large as 60\% of the spin phase is not very commonly observed from the known neutron stars. However, some magnetars and pulsars are known to exhibit emission over a large spin-phase range. For example, magnetar J1622$-$4950 exhibits radio emission covering slightly more than 60\% of the spin-phase \citep{Levin_2012}.
\magn is also known to show emission over as large as 30$-$35\% of the spin-period \citep[e.g., see][and from an assessment of the average profiles of new observations which are presented in this work]{Maan_2022}. Moreover, the average profile of \magn is known to change rapidly, sometimes even with emission components wandering around over spin-phase range of 60\% \citep[e.g., see][]{Camilo_2007}. Such changes have been observed over timescale of days and the faster timescales have not been probed systematically. \citet{Maan_2019} showed that the bright pulses from \magn typically cover the whole average emission region and these are not particularly limited to narrow phase-ranges.  We also note that the average profile observed on 1st October 2020 at 650\,MHz (corresponding to the observation used in Figure~\ref{fig:8}) happens to be narrow, with a duty cycle of the order of 4\%. The single pulses at this epoch sample a narrow phase range (top-right panel in Figure~\ref{fig:8}) decided by the average profile. At other epochs where the average profile is wide, the single pulses sample correspondingly large phase ranges.

The radio emission mechanism of magnetars in general, and FRB-like emission from magnetars in particular, is not well understood. Depending on the trigger mechanism (e.g., magnetic reconnection, star-quake induced magnetic twist or reconfigurations, etc.), the adequate physical conditions to initiate radio emission might be satisfied across a large fraction of the magnetosphere, and hence, over a large fraction of the spin phases. We also note that, apart from the bursts being emitted over a large fraction of the spin period, there could be other factors, e.g., binary motion, or precession of the source, that could make it difficult to detect any underlying periodicity.

\section{Conclusions} \label{sec-conclusion}
We have presented a detailed study of over 97000 low-energy bursts detected from the magnetar \magn, at a number of frequency bands covering 300 MHz to 6\,GHz frequency range and over a time span of 4.5\,years. The primary conclusions from our study are the following.
\begin{enumerate}
\item Magnetar \magn exhibits a variety of emission states, from normal pulsar-like emission to highly active giant-pulse emitting states. Irrespective of the observing frequency, the magnetar was found to be in pulsar-like emission states for a major fraction of the observing epochs, with burst fluences characterized by a lognormal distribution.
\item In favorable, giant-pulse emitting states, \magn could emit a burst as energetic as the Galactic FRB over a timescale of much less than a year. Given their dynamic emission behavior, magnetars emitting giant-pulses are much more likely to be the source of FRBs than Crab-like pulsars.
\item Detection of an underlying periodicity in the bursts from repeating FRBs might be hindered due to their emission over a wide range of spin-phases. We show that even a spread over 60\% of the spin period could make the underlying periodicity undetectable.
\end{enumerate}
Overall, the Galactic magnetars like \magn remain viable and likely sources of FRBs.

\acknowledgements
Acknowledgment: We would like to thank Tirthankar Roy Choudhury for his insightful discussions on model-fitting. BL acknowledges Akanksha Kapahtia for her valuable guidance on MCMC fitting and extends gratitude to Ajay Kumar and Yash Bhusare for their constructive discussions throughout the course of this project.
YM acknowledges support from the Department of Science and Technology via the Science and Engineering Research Board Startup Research Grant (SRG/2023/002657).
MA is supported by a graduate research assistantship funded by Tamkeen. JG is supported by NYU Abu Dhabi research grant AD 022.  Basic research at NYU Abu Dhabi is funded by the Executive Affairs Authority - Abu Dhabi, as administered by Tamkeen.
We would like to thank the Centre Director and the observatory for the prompt time allocation and scheduling of our observations. GMRT is run by the National Centre for Radio Astrophysics of the Tata Institute of Fundamental Research. We acknowledge the Department of Atomic Energy for funding support, under project 12$-$R\&D$-$TFR$-$5.02$-$0700.
The National Radio Astronomy Observatory is a facility of the National Science Foundation operated under cooperative agreement by Associated Universities, Inc.

Facility: GMRT(GWB) and GBT.

Software: RFIClean \citep{Maan_2021_rficlean}, PRESTO \citep{Ransom_2001}, SIGPROC, DSPSR \citep{dspsr_2011}, PSRCHIVE \citep{PSRCHIVE_2004}.

\bibliography{ref}{}
\bibliographystyle{aasjournal}

\appendix

\section{Box Plot} \label{box_plot}

\begin{figure*}[htp]
\centering
\includegraphics[scale=0.4]{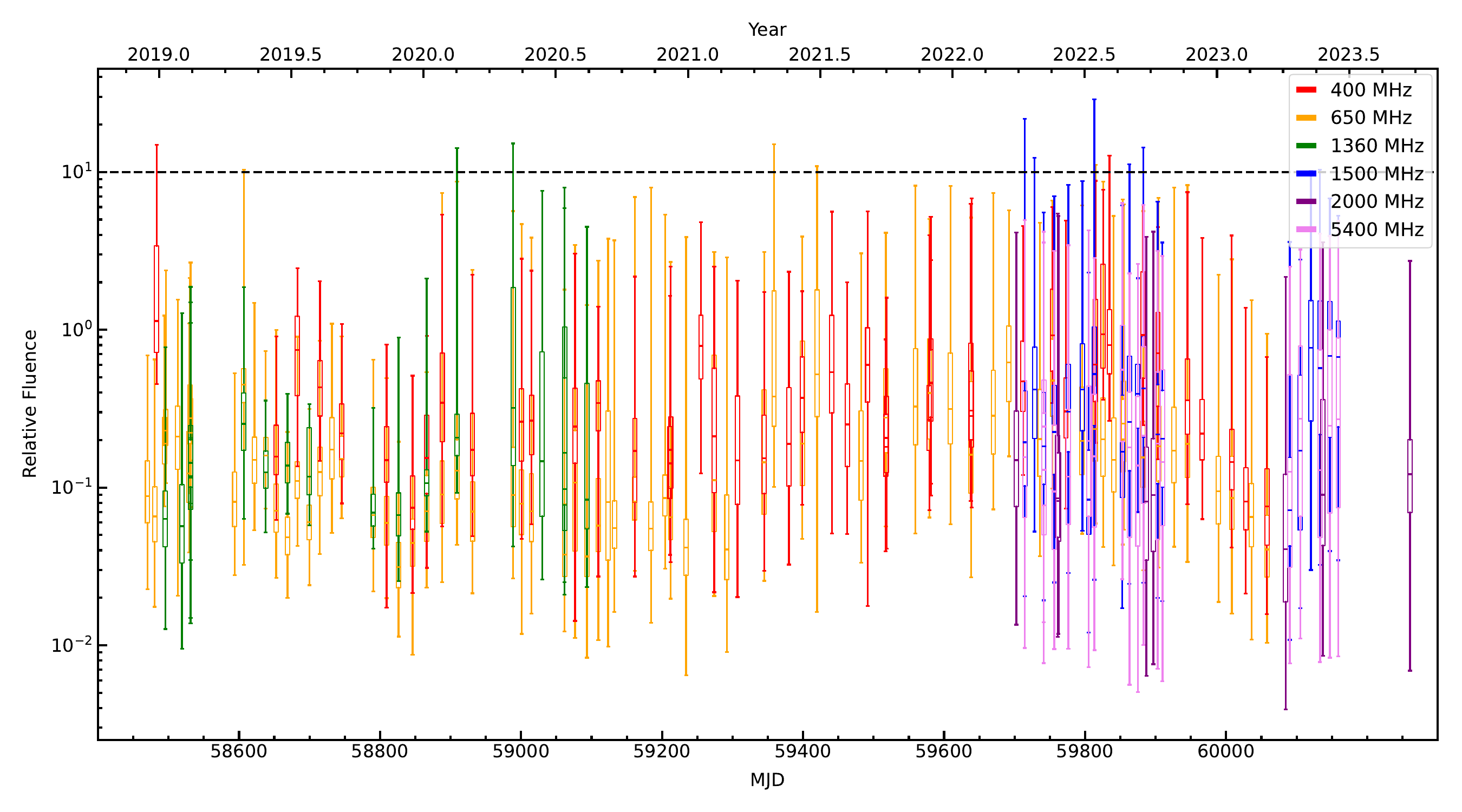}
\caption{The relative fluence distributions for all the observations using box plots, same as shown in Figure~\ref{fig:4} but with a log scale.}
\label{box_plot_log_scale}
\end{figure*}

\section{Fit for the deviation at distribution tail} \label{sec-appendix_A}
\begin{figure*}[htb]
    \centering
    \hspace*{-0.8cm}
    \includegraphics[scale=0.3]{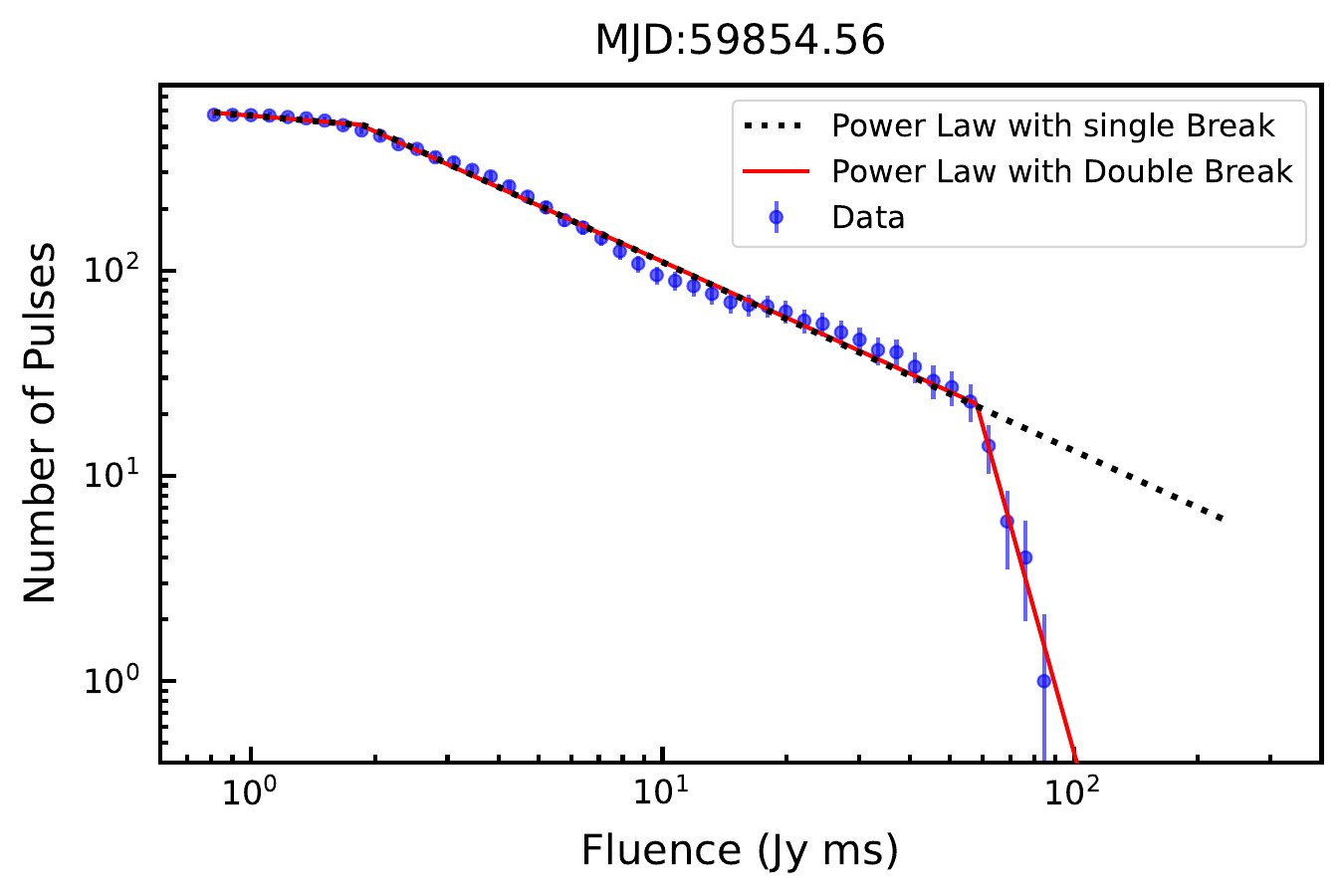}
    \caption{The cumulative fluence distribution fitted with a broken power-law (dotted black) and a power-law with two breaks (red).}
    \label{fig:9}
\end{figure*}

As mentioned in Section\ref{sec-distributions}, we have fitted the cumulative fluence distribution using four models. In a few cases where the best model was found to be the power-law or power-law with a break, although the fit is good (reduced chi-square is close to one), the tail-end apparently deviates from the fitted model (see, for example, Figure~\ref{fig:9}). For such cases, we have fitted the power-law with an additional break, as shown in Figure~\ref{fig:9} (red line). The timescale for a fluence of 1.5\,M\jyms is then computed using the rightmost power-law parameters. 

\section{Timescale from power law for the tail end} \label{sec-appendix_B}

\begin{figure*}
    \centering
    \includegraphics[scale=0.38]{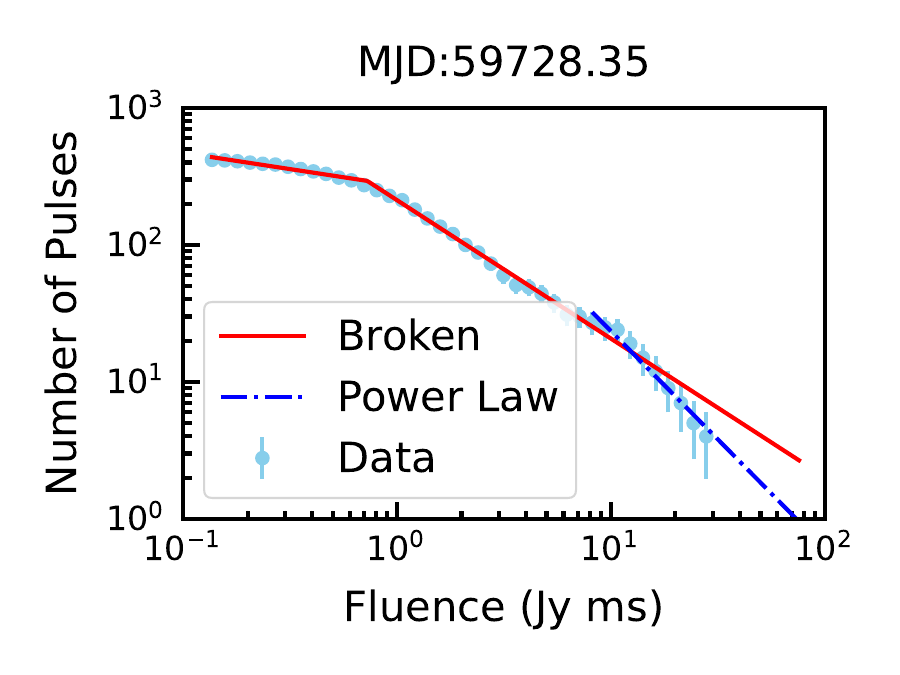}
    \hfill
    \includegraphics[scale=0.38]{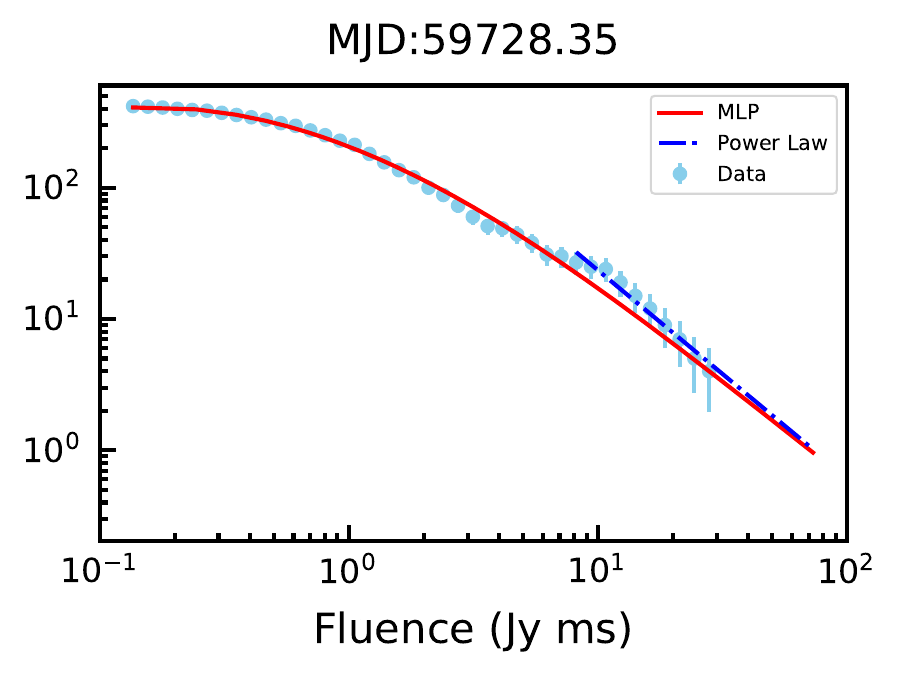}
    \hfill
    \includegraphics[scale=0.38]{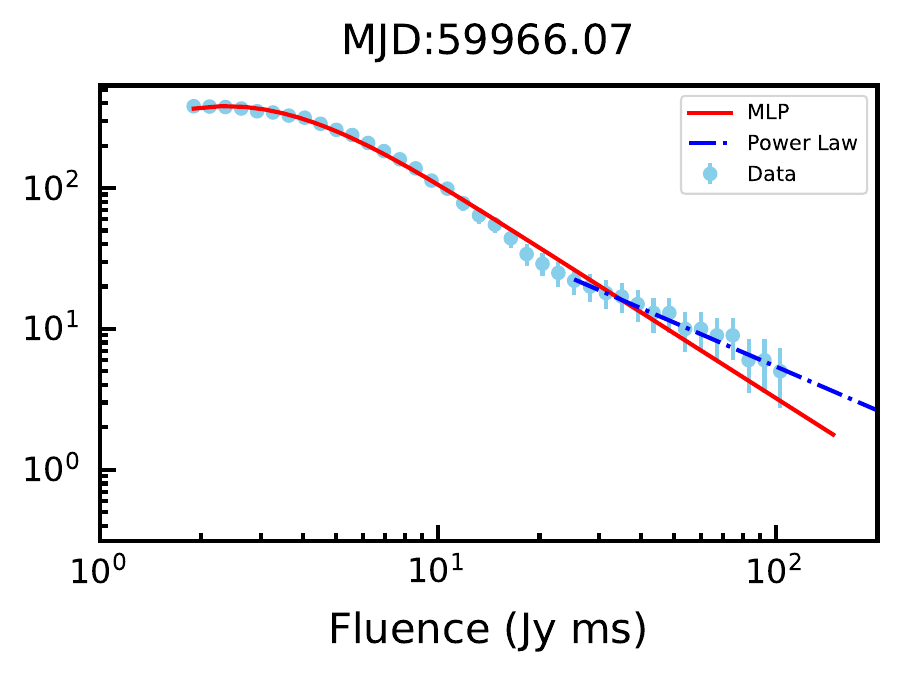}
    \caption{Cumulative fluence distributions at three different epochs are shown with the best-fitted model (red) as well as the power law characterization of only the tail-end (blue).}
    \label{fig:10}
\end{figure*}

As discussed in Section\ref{sec-timescale}, many distributions suggest the predicted timescale for a 1.5\,M\jyms to be less than 100 years. We have further scrutinized all these cases by characterizing their tail-ends using a power-law. For this characterization, first we chose the last 10 bins from the distribution and fit them with a power law, and computed the reduced chi-square value. Then we increase the number of chosen bins at the tail-end by one and again fit. This iteration is repeated until the last 20 bins are chosen. We then chose the best power-law fit based on the best reduced chi-square value. Some examples with the tail-end fitting are shown in figure~\ref{fig:10}.

\begin{figure*}[htp]
\centering
\includegraphics[width=\linewidth]{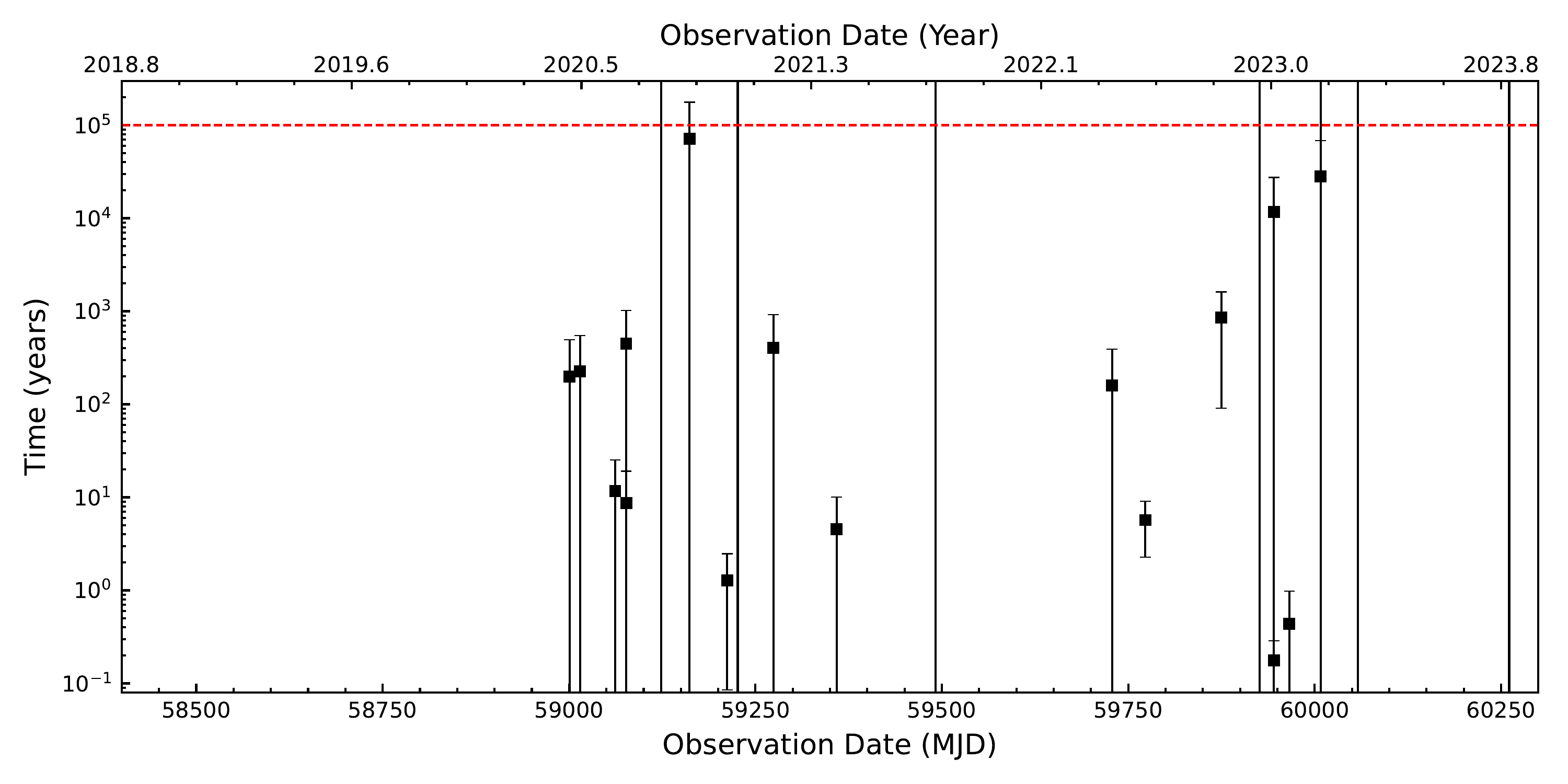}
\caption{Expected timescale derived from the power law characterization of only the tail-end, for a fluence of 1.5\,M\jyms is shown as a function of the observation date.}
\label{fig:11}
\end{figure*}

We notice that while comparing with the best-fitted model, the tail-end fitted power law is sometimes steeper (right panel in Figure~\ref{fig:10}), or similar to that from the original fit (middle panel), and sometimes flatter (left panel). We have also computed the timescale for the fluence 1.5\,M\jyms using the best-fitted power-law parameters. The timescale is shown in Figure~\ref{fig:11}, and overall the minimum timescale is similar to the minimum timescale in Figure~\ref{fig:6}.

\section{Example Corner Plots} \label{sec-appendix_C}
\begin{figure*}[htp]
\centering
\includegraphics[scale=0.35]{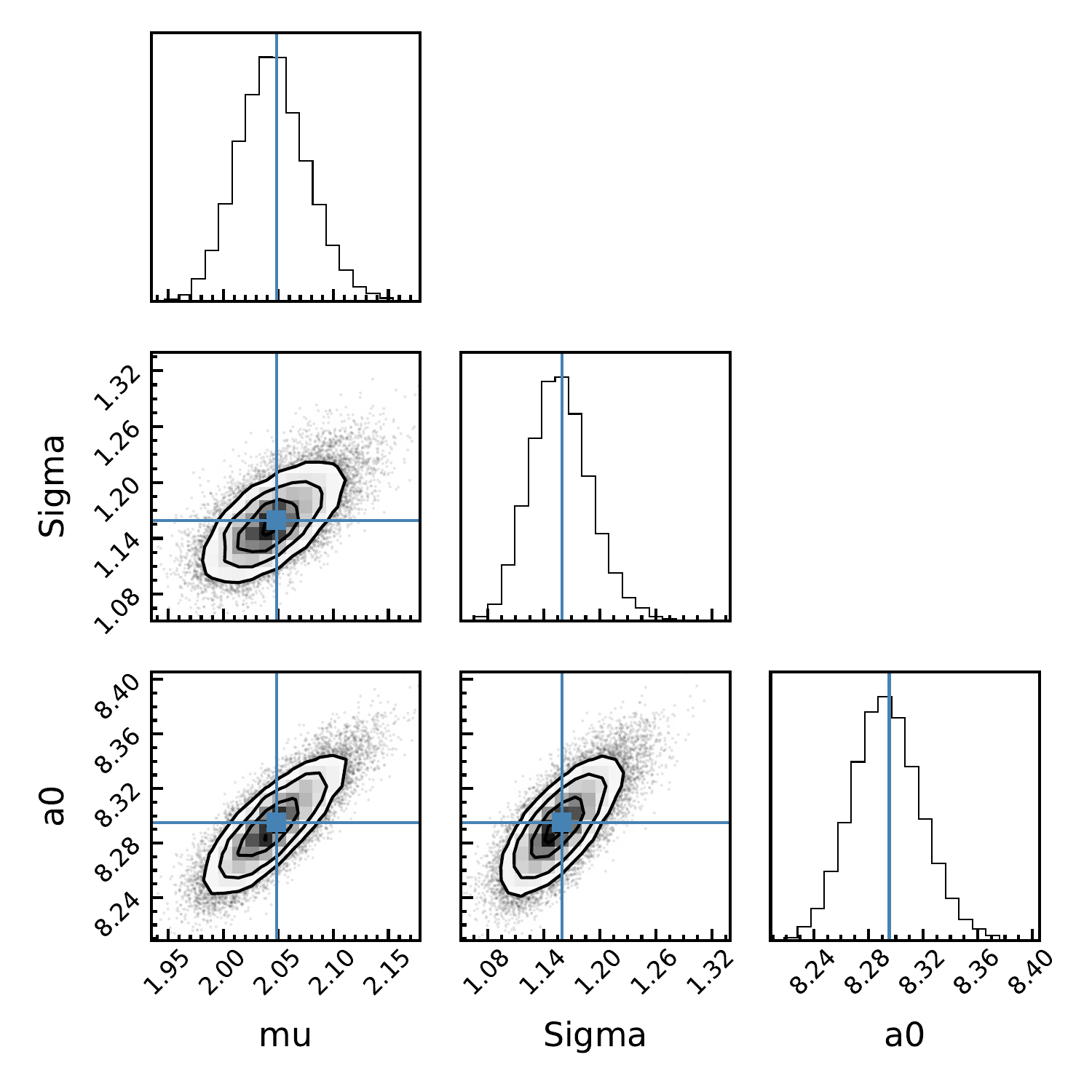}
\includegraphics[scale=0.35]{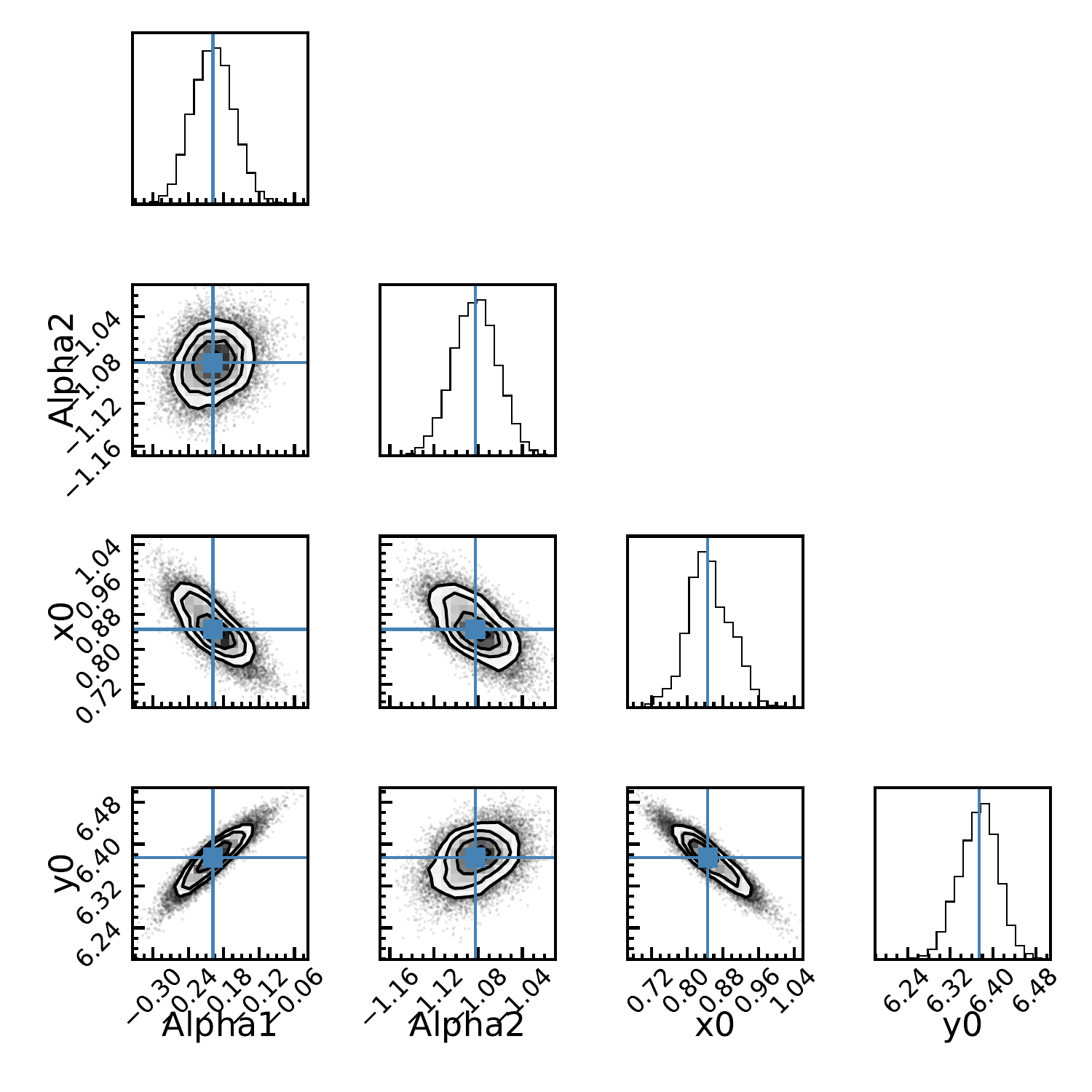}
\includegraphics[scale=0.35]{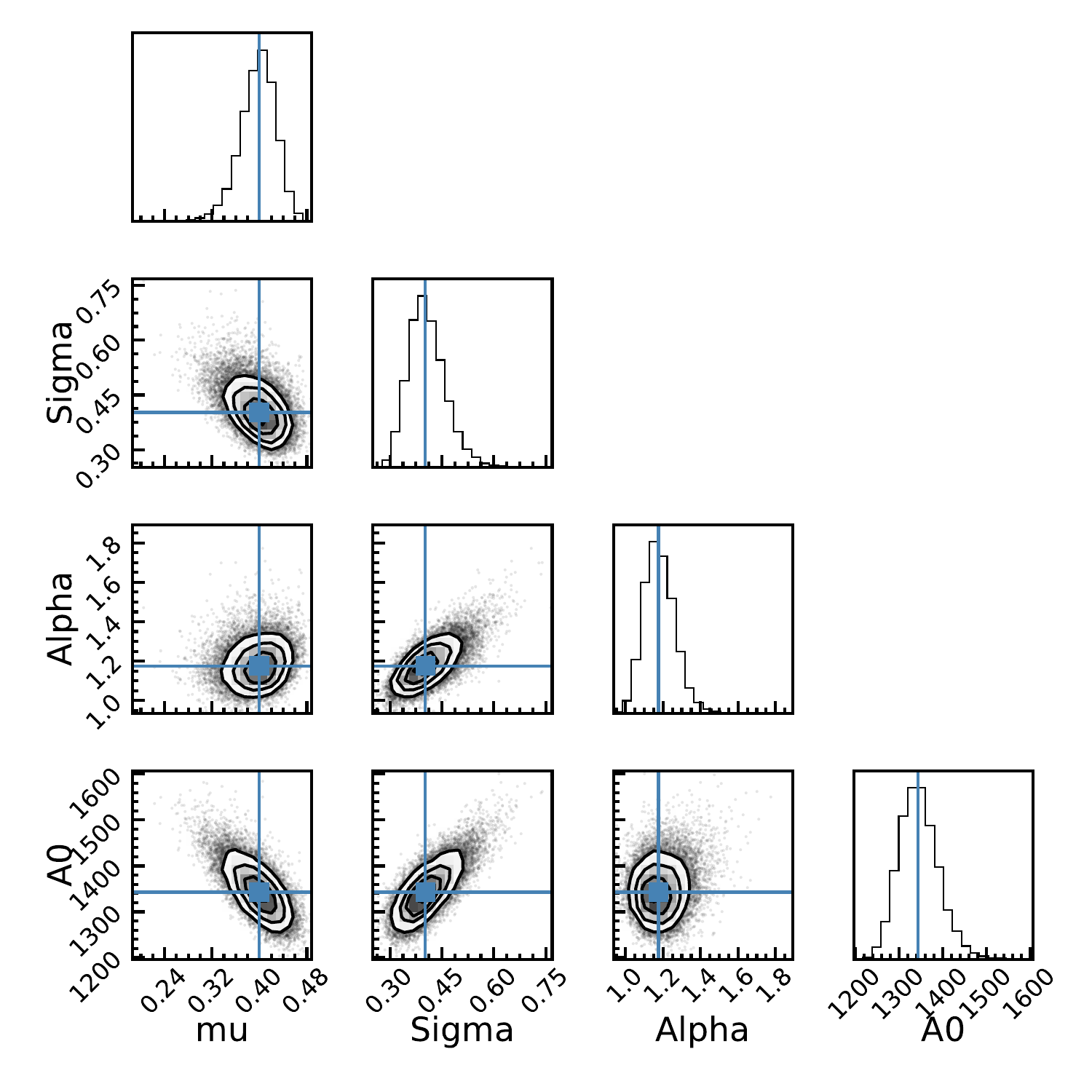}
\caption{Corner plots for fits at three different epochs with different models (from left: lognormal, broken power law, and MLP).}
\label{fig:12}
\end{figure*}

As mentioned in Section\ref{sec-distributions}, MCMC fitting provides us the posterior distributions for the model parameters. As shown by some examples in Figure~\ref{fig:12}, the model parameters are very well-constrained. 

\clearpage



\section{Table for goodness of fits} \label{sec-appendix_table}

The AICc values and reduced chi-square values are listed for each of the four models, for all the observations taken at different frequencies. The blue color (in bold) represents the values corresponding to the best model for the given observations, and the header information is as follows.

\setlength{\tabcolsep}{12pt}

\vspace{0.5em}
{\footnotesize
\noindent $\mathrm{aicc}_p$ = AICC value for power law.\\
\noindent $\mathrm{aicc}_b$ = AICC value for broken power law.\\
\noindent $\mathrm{aicc}_l$ = AICC value for lognormal.\\
\noindent $\mathrm{aicc}_m$ = AICC value for MLP.\\
$\chi{^2}_{re_{p}}$ = reduced chi-square for power law. \\
$\chi{^2}_{re_{b}}$ = reduced chi-square for broken power law. \\
$\chi{^2}_{re_{l}}$ = reduced chi-square for lognormal. \\
$\chi{^2}_{re_{m}}$ = reduced chi-square for MLP. \\
}

{\centering

\begin{longtable}{rllllllll}
\caption{At 400 MHz (band-3)} \\
\toprule
MJD & $aicc_p$ & $aicc_b$ & $aicc_l$ & $aicc_m$ & $\chi{^2}_{re_{p}}$ & $\chi{^2}_{re_{b}}$ & $\chi{^2}_{re_{l}}$ & $\chi{^2}_{re_{m}}$ \\
\midrule
\endfirsthead
\toprule
MJD & $aicc_p$ & $aicc_b$ & $aicc_l$ & $aicc_m$ & $\chi{^2}_{re_{p}}$ & $\chi{^2}_{re_{b}}$ & $\chi{^2}_{re_{l}}$ & $\chi{^2}_{re_{m}}$ \\
\midrule
\endhead
\midrule
\multicolumn{9}{r}{Continued on next page} \\
\midrule
\endfoot
\bottomrule
\endlastfoot
58483.1 & \textcolor{blue}{\textbf{8.4}} & 15.7 & 18.4 & 16.6 & \textcolor{blue}{\textbf{0.3}} & 0.2 & 1.0 & 0.4 \\
58652.8 & \textcolor{blue}{\textbf{10.6}} & 20.9 & 13.2 & 17.5 & \textcolor{blue}{\textbf{0.7}} & 0.8 & 0.5 & 0.2 \\
58682.8 & 37.0 & 14.8 & \textcolor{blue}{\textbf{13.8}} & 18.0 & 2.5 & 0.3 & \textcolor{blue}{\textbf{0.4}} & 0.5 \\
58714.8 & 47.7 & 18.2 & \textcolor{blue}{\textbf{11.8}} & 16.0 & 2.9 & 0.5 & \textcolor{blue}{\textbf{0.3}} & 0.4 \\
58745.7 & 49.6 & 15.2 & \textcolor{blue}{\textbf{8.7}} & 12.1 & 3.0 & 0.3 & \textcolor{blue}{\textbf{0.1}} & 0.1 \\
58809.2 & 312.8 & \textcolor{blue}{\textbf{38.9}} & 47.2 & 60.0 & 13.4 & \textcolor{blue}{\textbf{1.3}} & 1.8 & 2.3 \\
58846.2 & 560.1 & 56.7 & \textcolor{blue}{\textbf{33.2}} & 37.9 & 14.2 & 1.2 & \textcolor{blue}{\textbf{0.7}} & 0.8 \\
58866.2 & 310.2 & 66.7 & \textcolor{blue}{\textbf{30.3}} & 57.5 & 9.3 & 1.8 & \textcolor{blue}{\textbf{0.7}} & 1.5 \\
58888.1 & 202.0 & 59.4 & \textcolor{blue}{\textbf{21.7}} & 26.6 & 7.3 & 1.9 & \textcolor{blue}{\textbf{0.6}} & 0.6 \\
58931.0 & 217.5 & 30.6 & \textcolor{blue}{\textbf{23.3}} & 25.1 & 7.9 & 0.8 & \textcolor{blue}{\textbf{0.6}} & 0.6 \\
59000.7 & 905.2 & 67.3 & \textcolor{blue}{\textbf{52.0}} & 56.8 & 21.0 & 1.4 & \textcolor{blue}{\textbf{1.1}} & 1.1 \\
59014.8 & 82.2 & \textcolor{blue}{\textbf{12.8}} & 13.6 & 16.9 & 4.6 & \textcolor{blue}{\textbf{0.1}} & 0.4 & 0.4 \\
59076.6 & 959.8 & \textcolor{blue}{\textbf{52.8}} & 4860.0 & 71.5 & 25.8 & \textcolor{blue}{\textbf{1.2}} & 131.2 & 1.7 \\
59109.5 & 6205.0 & \textcolor{blue}{\textbf{393.7}} & 4793.9 & 2638.2 & 87.3 & \textcolor{blue}{\textbf{5.5}} & 67.4 & 37.6 \\
59161.5 & 691.2 & \textcolor{blue}{\textbf{34.0}} & 48.6 & 51.7 & 16.8 & \textcolor{blue}{\textbf{0.6}} & 1.0 & 1.1 \\
59211.3 & 156.3 & 28.8 & \textcolor{blue}{\textbf{16.9}} & 29.9 & 5.6 & 0.7 & \textcolor{blue}{\textbf{0.4}} & 0.8 \\
59212.2 & 292.2 & 34.9 & \textcolor{blue}{\textbf{11.5}} & 14.4 & 9.3 & 0.8 & \textcolor{blue}{\textbf{0.2}} & 0.2 \\
59255.1 & 564.7 & \textcolor{blue}{\textbf{54.8}} & 77.7 & 111.6 & 17.0 & \textcolor{blue}{\textbf{1.4}} & 2.2 & 3.2 \\
59274.0 & 170.8 & \textcolor{blue}{\textbf{56.1}} & 458.2 & 93.1 & 6.2 & \textcolor{blue}{\textbf{1.8}} & 16.7 & 3.2 \\
59307.0 & 541.0 & 145.6 & 405.8 & \textcolor{blue}{\textbf{139.3}} & 12.5 & 3.2 & 9.3 & \textcolor{blue}{\textbf{3.1}} \\
59344.8 & 321.8 & 65.4 & \textcolor{blue}{\textbf{28.0}} & 43.4 & 9.1 & 1.6 & \textcolor{blue}{\textbf{0.6}} & 1.0 \\
59380.0 & 312.5 & 88.8 & 65.6 & \textcolor{blue}{\textbf{60.8}} & 8.3 & 2.2 & 1.6 & \textcolor{blue}{\textbf{1.4}} \\
59398.8 & 87.3 & 24.1 & \textcolor{blue}{\textbf{14.0}} & 17.9 & 4.4 & 0.8 & \textcolor{blue}{\textbf{0.4}} & 0.4 \\
59440.8 & 551.8 & 103.9 & 118.6 & \textcolor{blue}{\textbf{86.4}} & 15.6 & 2.8 & 3.2 & \textcolor{blue}{\textbf{2.3}} \\
59462.6 & 231.6 & 30.5 & \textcolor{blue}{\textbf{14.8}} & 18.5 & 7.3 & 0.7 & \textcolor{blue}{\textbf{0.3}} & 0.3 \\
59491.6 & 847.2 & \textcolor{blue}{\textbf{70.7}} & 158.6 & 200.6 & 25.5 & \textcolor{blue}{\textbf{1.9}} & 4.6 & 6.0 \\
59517.3 & 147.6 & 20.8 & \textcolor{blue}{\textbf{12.4}} & 15.4 & 6.2 & 0.5 & \textcolor{blue}{\textbf{0.2}} & 0.2 \\
59518.3 & 363.2 & 52.0 & \textcolor{blue}{\textbf{12.9}} & 15.5 & 10.9 & 1.3 & \textcolor{blue}{\textbf{0.2}} & 0.2 \\
59579.1 & 184.5 & 27.7 & \textcolor{blue}{\textbf{10.1}} & 12.0 & 6.2 & 0.6 & \textcolor{blue}{\textbf{0.1}} & 0.1 \\
59581.3 & 40.4 & 15.4 & \textcolor{blue}{\textbf{9.7}} & 13.7 & 2.4 & 0.3 & \textcolor{blue}{\textbf{0.1}} & 0.2 \\
59581.2 & 203.9 & 29.9 & \textcolor{blue}{\textbf{12.2}} & 14.9 & 6.9 & 0.7 & \textcolor{blue}{\textbf{0.2}} & 0.2 \\
59638.0 & 41.9 & 31.7 & 26.6 & \textcolor{blue}{\textbf{26.1}} & 1.6 & 1.0 & 0.8 & \textcolor{blue}{\textbf{0.7}} \\
59638.9 & 127.3 & 27.2 & \textcolor{blue}{\textbf{22.9}} & 23.9 & 4.0 & 0.6 & \textcolor{blue}{\textbf{0.5}} & 0.5 \\
59711.8 & 125.3 & 31.2 & \textcolor{blue}{\textbf{24.5}} & 25.0 & 4.5 & 0.8 & \textcolor{blue}{\textbf{0.6}} & 0.6 \\
59753.8 & \textcolor{blue}{\textbf{8.4}} & 13.7 & 9.5 & 13.0 & \textcolor{blue}{\textbf{0.3}} & 0.2 & 0.1 & 0.1 \\
59772.8 & 142.2 & \textcolor{blue}{\textbf{14.5}} & 33.6 & 16.7 & 5.5 & \textcolor{blue}{\textbf{0.2}} & 1.1 & 0.3 \\
59814.5 & 18.6 & \textcolor{blue}{\textbf{13.4}} & 15.8 & 19.5 & 0.7 & \textcolor{blue}{\textbf{0.2}} & 0.4 & 0.5 \\
59825.6 & \textcolor{blue}{\textbf{7.0}} & 12.6 & 9.8 & 13.7 & \textcolor{blue}{\textbf{0.2}} & 0.0 & 0.1 & 0.1 \\
59834.6 & 23.7 & 14.0 & \textcolor{blue}{\textbf{13.3}} & 14.3 & 1.1 & 0.2 & \textcolor{blue}{\textbf{0.3}} & 0.2 \\
59882.3 & 23.9 & 15.3 & \textcolor{blue}{\textbf{12.5}} & 17.7 & 1.3 & 0.3 & \textcolor{blue}{\textbf{0.3}} & 0.5 \\
59903.3 & 18.5 & 14.1 & \textcolor{blue}{\textbf{9.7}} & 13.0 & 0.9 & 0.2 & \textcolor{blue}{\textbf{0.1}} & 0.1 \\
59945.2 & 139.0 & \textcolor{blue}{\textbf{23.8}} & 37.9 & 38.4 & 4.3 & \textcolor{blue}{\textbf{0.5}} & 1.0 & 1.0 \\
59966.1 & 276.5 & \textcolor{blue}{\textbf{28.5}} & 92.3 & 30.4 & 7.8 & \textcolor{blue}{\textbf{0.6}} & 2.4 & 0.6 \\
60008.0 & 417.9 & \textcolor{blue}{\textbf{40.4}} & 134.6 & 50.1 & 9.2 & \textcolor{blue}{\textbf{0.7}} & 2.8 & 0.9 \\
60027.9 & 1126.5 & 87.4 & 107.3 & \textcolor{blue}{\textbf{50.5}} & 21.2 & 1.5 & 1.9 & \textcolor{blue}{\textbf{0.8}} \\
60058.0 & 1532.6 & 115.5 & \textcolor{blue}{\textbf{85.3}} & 110.0 & 27.8 & 2.0 & \textcolor{blue}{\textbf{1.4}} & 1.9 \\
\end{longtable}

\par}

{\centering
\begin{longtable}{rrlllrlll}
\caption{At 650 MHz (band-4)} \\
\toprule
MJD & $aicc_p$ & $aicc_b$ & $aicc_l$ & $aicc_m$ & $\chi{^2}_{re_{p}}$ & $\chi{^2}_{re_{b}}$ & $\chi{^2}_{re_{l}}$ & $\chi{^2}_{re_{m}}$ \\
\midrule
\endfirsthead
\toprule
MJD & $aicc_p$ & $aicc_b$ & $aicc_l$ & $aicc_m$ & $\chi{^2}_{re_{p}}$ & $\chi{^2}_{re_{b}}$ & $\chi{^2}_{re_{l}}$ & $\chi{^2}_{re_{m}}$ \\
\midrule
\endhead
\midrule
\multicolumn{9}{r}{Continued on next page} \\
\midrule
\endfoot
\bottomrule
\endlastfoot
58470.2 & 1553.4 & 227.8 & \textcolor{blue}{\textbf{38.0}} & 55.8 & 27.2 & 3.9 & \textcolor{blue}{\textbf{0.6}} & 0.8 \\
58480.4 & 5631.5 & 490.2 & \textcolor{blue}{\textbf{22.3}} & 23.6 & 66.2 & 5.7 & \textcolor{blue}{\textbf{0.2}} & 0.2 \\
58494.3 & 409.1 & 60.0 & 20.4 & \textcolor{blue}{\textbf{16.5}} & 10.4 & 1.3 & 0.4 & \textcolor{blue}{\textbf{0.2}} \\
58496.3 & 191.8 & \textcolor{blue}{\textbf{23.8}} & 65.8 & 27.0 & 5.4 & \textcolor{blue}{\textbf{0.4}} & 1.7 & 0.5 \\
58513.0 & 708.2 & 82.4 & \textcolor{blue}{\textbf{78.6}} & 99.2 & 21.3 & 2.3 & \textcolor{blue}{\textbf{2.2}} & 2.8 \\
58529.1 & 626.8 & 164.1 & \textcolor{blue}{\textbf{48.6}} & 61.3 & 12.2 & 3.1 & \textcolor{blue}{\textbf{0.8}} & 1.0 \\
58530.2 & 543.9 & 70.5 & \textcolor{blue}{\textbf{20.4}} & 23.4 & 13.2 & 1.5 & \textcolor{blue}{\textbf{0.3}} & 0.4 \\
58531.2 & 1221.2 & 235.4 & \textcolor{blue}{\textbf{48.8}} & 51.3 & 19.3 & 3.7 & \textcolor{blue}{\textbf{0.7}} & 0.7 \\
58593.8 & 438.2 & 57.8 & \textcolor{blue}{\textbf{11.2}} & 15.1 & 11.1 & 1.3 & \textcolor{blue}{\textbf{0.1}} & 0.2 \\
58606.8 & 229.6 & \textcolor{blue}{\textbf{31.0}} & 70.7 & 76.9 & 11.8 & \textcolor{blue}{\textbf{1.2}} & 3.3 & 3.7 \\
58621.8 & 439.9 & 51.0 & 40.0 & \textcolor{blue}{\textbf{25.4}} & 11.8 & 1.2 & 0.9 & \textcolor{blue}{\textbf{0.4}} \\
58637.9 & 357.7 & 38.8 & \textcolor{blue}{\textbf{14.3}} & 15.0 & 10.1 & 0.9 & \textcolor{blue}{\textbf{0.2}} & 0.2 \\
58652.8 & 1510.2 & 148.0 & 46.9 & \textcolor{blue}{\textbf{20.5}} & 23.9 & 2.2 & 0.6 & \textcolor{blue}{\textbf{0.2}} \\
58668.8 & 2881.1 & 277.3 & \textcolor{blue}{\textbf{21.4}} & 27.4 & 41.7 & 4.0 & \textcolor{blue}{\textbf{0.2}} & 0.3 \\
58682.7 & 1640.0 & 83.7 & 75.9 & \textcolor{blue}{\textbf{35.5}} & 29.7 & 1.4 & 1.3 & \textcolor{blue}{\textbf{0.5}} \\
58699.7 & 3212.0 & 188.4 & 36.5 & \textcolor{blue}{\textbf{33.9}} & 46.5 & 2.6 & 0.4 & \textcolor{blue}{\textbf{0.4}} \\
58714.7 & 1564.0 & 179.1 & \textcolor{blue}{\textbf{18.2}} & 21.4 & 28.4 & 3.2 & \textcolor{blue}{\textbf{0.2}} & 0.2 \\
58731.7 & 868.8 & 107.5 & \textcolor{blue}{\textbf{31.6}} & 54.6 & 18.4 & 2.1 & \textcolor{blue}{\textbf{0.5}} & 1.0 \\
58745.6 & 267.8 & 26.3 & 18.9 & \textcolor{blue}{\textbf{18.7}} & 8.0 & 0.5 & 0.4 & \textcolor{blue}{\textbf{0.3}} \\
58790.4 & 1524.7 & 155.0 & \textcolor{blue}{\textbf{14.7}} & 15.6 & 26.7 & 2.6 & \textcolor{blue}{\textbf{0.1}} & 0.1 \\
58809.3 & 3291.4 & 286.9 & \textcolor{blue}{\textbf{22.2}} & 26.9 & 43.8 & 3.8 & \textcolor{blue}{\textbf{0.2}} & 0.2 \\
58826.3 & 3935.7 & 276.0 & 51.5 & \textcolor{blue}{\textbf{21.2}} & 49.8 & 3.4 & 0.6 & \textcolor{blue}{\textbf{0.2}} \\
58846.2 & 8926.3 & 532.5 & \textcolor{blue}{\textbf{375.1}} & 386.1 & 102.6 & 6.1 & \textcolor{blue}{\textbf{4.2}} & 4.4 \\
58866.2 & 240.5 & 55.6 & \textcolor{blue}{\textbf{19.9}} & 25.3 & 7.2 & 1.4 & \textcolor{blue}{\textbf{0.4}} & 0.5 \\
58888.0 & 1246.5 & 151.1 & 600.7 & \textcolor{blue}{\textbf{104.3}} & 19.1 & 2.2 & 9.1 & \textcolor{blue}{\textbf{1.5}} \\
58909.1 & 574.6 & 68.4 & 196.6 & \textcolor{blue}{\textbf{29.4}} & 10.8 & 1.2 & 3.6 & \textcolor{blue}{\textbf{0.4}} \\
58931.1 & 1110.2 & 71.7 & 236.8 & \textcolor{blue}{\textbf{42.2}} & 18.7 & 1.1 & 3.9 & \textcolor{blue}{\textbf{0.6}} \\
58988.8 & 115.7 & \textcolor{blue}{\textbf{89.6}} & 565.3 & 150.5 & 2.7 & \textcolor{blue}{\textbf{2.0}} & 13.6 & 3.5 \\
59000.8 & 2371.7 & \textcolor{blue}{\textbf{70.0}} & 365.4 & 97.3 & 38.8 & \textcolor{blue}{\textbf{1.0}} & 5.9 & 1.5 \\
59014.7 & 1279.5 & 48.0 & 191.1 & \textcolor{blue}{\textbf{30.4}} & 20.2 & 0.6 & 2.9 & \textcolor{blue}{\textbf{0.4}} \\
59029.9 & 274.0 & \textcolor{blue}{\textbf{191.6}} & 556.5 & 311.8 & 5.1 & \textcolor{blue}{\textbf{3.5}} & 10.4 & 5.8 \\
59061.8 & 1207.6 & \textcolor{blue}{\textbf{845.0}} & 1618.3 & 2381.7 & 13.5 & \textcolor{blue}{\textbf{9.5}} & 18.1 & 27.0 \\
59076.6 & 543.5 & 39.6 & 889.5 & \textcolor{blue}{\textbf{35.6}} & 12.0 & 0.7 & 19.6 & \textcolor{blue}{\textbf{0.6}} \\
59093.4 & 5674.4 & \textcolor{blue}{\textbf{659.8}} & 2703.1 & 1132.5 & 57.3 & \textcolor{blue}{\textbf{6.6}} & 27.2 & 11.5 \\
59109.6 & 8050.4 & \textcolor{blue}{\textbf{280.2}} & 1587.1 & 721.8 & 70.0 & \textcolor{blue}{\textbf{2.4}} & 13.8 & 6.3 \\
59123.5 & 694.0 & \textcolor{blue}{\textbf{224.0}} & 4636.1 & 545.3 & 11.3 & \textcolor{blue}{\textbf{3.6}} & 75.9 & 8.9 \\
59132.5 & 1991.3 & 1954.8 & 1715.1 & \textcolor{blue}{\textbf{602.2}} & 25.2 & 25.0 & 21.6 & \textcolor{blue}{\textbf{7.6}} \\
59161.4 & 479.7 & \textcolor{blue}{\textbf{279.8}} & 552.3 & 449.2 & 10.6 & \textcolor{blue}{\textbf{6.2}} & 12.1 & 10.0 \\
59184.4 & 1328.4 & 1331.0 & 1150.2 & \textcolor{blue}{\textbf{377.5}} & 21.0 & 21.3 & 18.2 & \textcolor{blue}{\textbf{6.0}} \\
59204.3 & 307.5 & 310.5 & 288.3 & \textcolor{blue}{\textbf{112.2}} & 6.7 & 6.8 & 6.3 & \textcolor{blue}{\textbf{2.4}} \\
59212.1 & 244.5 & \textcolor{blue}{\textbf{24.2}} & 115.6 & 26.1 & 6.5 & \textcolor{blue}{\textbf{0.4}} & 2.9 & 0.5 \\
59234.1 & 5252.4 & \textcolor{blue}{\textbf{600.7}} & 2307.2 & 679.7 & 64.8 & \textcolor{blue}{\textbf{7.4}} & 28.4 & 8.4 \\
59274.0 & 82.9 & \textcolor{blue}{\textbf{22.4}} & 203.9 & 94.4 & 2.9 & \textcolor{blue}{\textbf{0.5}} & 7.3 & 3.3 \\
59291.9 & 864.4 & \textcolor{blue}{\textbf{557.5}} & 577.3 & 760.2 & 11.5 & \textcolor{blue}{\textbf{7.4}} & 7.6 & 10.2 \\
59344.9 & 224.1 & 61.7 & \textcolor{blue}{\textbf{39.9}} & 82.6 & 5.9 & 1.5 & \textcolor{blue}{\textbf{0.9}} & 2.0 \\
59358.8 & 45.6 & \textcolor{blue}{\textbf{23.6}} & 68.4 & 47.5 & 1.5 & \textcolor{blue}{\textbf{0.5}} & 2.3 & 1.5 \\
59398.9 & 104.4 & \textcolor{blue}{\textbf{21.9}} & 79.9 & 87.5 & 3.2 & \textcolor{blue}{\textbf{0.4}} & 2.4 & 2.6 \\
59419.8 & 1570.9 & 232.6 & \textcolor{blue}{\textbf{166.4}} & 385.7 & 32.0 & 4.7 & \textcolor{blue}{\textbf{3.3}} & 7.8 \\
59462.7 & 46.6 & 21.8 & 31.1 & \textcolor{blue}{\textbf{14.4}} & 1.7 & 0.5 & 1.0 & \textcolor{blue}{\textbf{0.2}} \\
59482.4 & 954.6 & 281.0 & \textcolor{blue}{\textbf{114.9}} & 287.6 & 15.6 & 4.5 & \textcolor{blue}{\textbf{1.8}} & 4.6 \\
59517.4 & 536.0 & 116.8 & \textcolor{blue}{\textbf{37.6}} & 62.5 & 12.4 & 2.6 & \textcolor{blue}{\textbf{0.7}} & 1.3 \\
59559.4 & 1131.2 & 264.9 & \textcolor{blue}{\textbf{64.2}} & 130.7 & 20.5 & 4.7 & \textcolor{blue}{\textbf{1.0}} & 2.3 \\
59579.2 & 183.6 & 41.1 & \textcolor{blue}{\textbf{16.4}} & 27.0 & 6.2 & 1.1 & \textcolor{blue}{\textbf{0.3}} & 0.6 \\
59609.0 & 601.8 & 98.6 & \textcolor{blue}{\textbf{73.1}} & 95.7 & 11.7 & 1.8 & \textcolor{blue}{\textbf{1.3}} & 1.7 \\
59638.1 & 212.6 & 103.5 & \textcolor{blue}{\textbf{44.0}} & 52.2 & 5.6 & 2.6 & \textcolor{blue}{\textbf{1.0}} & 1.2 \\
59669.9 & 129.4 & 23.9 & \textcolor{blue}{\textbf{22.9}} & 25.2 & 3.8 & 0.5 & \textcolor{blue}{\textbf{0.5}} & 0.5 \\
59692.0 & 36.1 & 16.1 & \textcolor{blue}{\textbf{10.1}} & 13.3 & 2.1 & 0.4 & \textcolor{blue}{\textbf{0.2}} & 0.2 \\
59735.8 & 435.8 & 59.4 & \textcolor{blue}{\textbf{49.3}} & 62.5 & 10.0 & 1.2 & \textcolor{blue}{\textbf{1.0}} & 1.3 \\
59753.8 & 110.6 & 19.2 & \textcolor{blue}{\textbf{13.2}} & 14.4 & 4.2 & 0.4 & \textcolor{blue}{\textbf{0.2}} & 0.2 \\
59795.7 & 139.4 & \textcolor{blue}{\textbf{55.6}} & 95.7 & 109.7 & 3.3 & \textcolor{blue}{\textbf{1.2}} & 2.2 & 2.5 \\
59814.6 & 107.3 & \textcolor{blue}{\textbf{28.3}} & 58.7 & 29.5 & 3.1 & \textcolor{blue}{\textbf{0.6}} & 1.6 & 0.6 \\
59825.6 & 198.2 & \textcolor{blue}{\textbf{44.5}} & 94.4 & 47.1 & 5.5 & \textcolor{blue}{\textbf{1.0}} & 2.5 & 1.1 \\
59840.5 & 1776.7 & \textcolor{blue}{\textbf{158.8}} & 558.4 & 354.7 & 21.9 & \textcolor{blue}{\textbf{1.9}} & 6.8 & 4.3 \\
59853.5 & 1378.6 & \textcolor{blue}{\textbf{109.4}} & 389.3 & 303.2 & 19.9 & \textcolor{blue}{\textbf{1.5}} & 5.6 & 4.3 \\
59854.6 & 285.2 & \textcolor{blue}{\textbf{59.2}} & 122.2 & 137.2 & 6.5 & \textcolor{blue}{\textbf{1.2}} & 2.7 & 3.0 \\
59882.4 & 950.8 & \textcolor{blue}{\textbf{42.6}} & 225.8 & 83.1 & 17.9 & \textcolor{blue}{\textbf{0.6}} & 4.1 & 1.4 \\
59904.3 & 8116.2 & \textcolor{blue}{\textbf{398.4}} & 1290.5 & 1201.3 & 70.5 & \textcolor{blue}{\textbf{3.4}} & 11.2 & 10.5 \\
59926.2 & 317.0 & \textcolor{blue}{\textbf{37.5}} & 54.5 & 49.0 & 7.0 & \textcolor{blue}{\textbf{0.6}} & 1.1 & 0.9 \\
59945.2 & 739.5 & \textcolor{blue}{\textbf{46.6}} & 101.0 & 56.5 & 15.0 & \textcolor{blue}{\textbf{0.8}} & 1.9 & 1.0 \\
59989.1 & 1785.1 & 106.6 & 56.7 & \textcolor{blue}{\textbf{29.7}} & 30.2 & 1.7 & 0.8 & \textcolor{blue}{\textbf{0.4}} \\
60008.1 & 2110.7 & 123.1 & 145.4 & \textcolor{blue}{\textbf{45.3}} & 32.4 & 1.8 & 2.1 & \textcolor{blue}{\textbf{0.6}} \\
60036.0 & 4086.1 & \textcolor{blue}{\textbf{123.7}} & 356.9 & 126.9 & 55.9 & \textcolor{blue}{\textbf{1.6}} & 4.8 & 1.6 \\
60058.0 & 3318.1 & 114.8 & 368.2 & \textcolor{blue}{\textbf{68.8}} & 39.0 & 1.3 & 4.3 & \textcolor{blue}{\textbf{0.7}} \\
\end{longtable}

\par}

{\centering
\begin{longtable}{rllllllll}
\caption{At 1360 MHz (band-5)} \\
\toprule
MJD & $aicc_p$ & $aicc_b$ & $aicc_l$ & $aicc_m$ & $\chi{^2}_{re_{p}}$ & $\chi{^2}_{re_{b}}$ & $\chi{^2}_{re_{l}}$ & $\chi{^2}_{re_{m}}$ \\
\midrule
\endfirsthead
\toprule
MJD & $aicc_p$ & $aicc_b$ & $aicc_l$ & $aicc_m$ & $\chi{^2}_{re_{p}}$ & $\chi{^2}_{re_{b}}$ & $\chi{^2}_{re_{l}}$ & $\chi{^2}_{re_{m}}$ \\
\midrule
\endhead
\midrule
\multicolumn{9}{r}{Continued on next page} \\
\midrule
\endfoot
\bottomrule
\endlastfoot
58495.4 & 6948.7 & 374.6 & \textcolor{blue}{\textbf{100.6}} & 7862.5 & 81.7 & 4.4 & \textcolor{blue}{\textbf{1.1}} & 93.5 \\
58519.1 & 3082.2 & 308.3 & \textcolor{blue}{\textbf{20.9}} & 27.8 & 43.4 & 4.3 & \textcolor{blue}{\textbf{0.2}} & 0.3 \\
58531.2 & 3537.7 & 335.3 & \textcolor{blue}{\textbf{164.5}} & 254.2 & 56.1 & 5.3 & \textcolor{blue}{\textbf{2.5}} & 4.0 \\
58531.3 & 654.9 & 58.9 & \textcolor{blue}{\textbf{17.7}} & 18.2 & 15.9 & 1.2 & \textcolor{blue}{\textbf{0.3}} & 0.2 \\
58531.2 & 3880.5 & 349.6 & \textcolor{blue}{\textbf{97.5}} & 165.3 & 57.9 & 5.2 & \textcolor{blue}{\textbf{1.4}} & 2.4 \\
58606.8 & 1060.4 & 107.7 & \textcolor{blue}{\textbf{18.9}} & 23.7 & 22.5 & 2.2 & \textcolor{blue}{\textbf{0.3}} & 0.3 \\
58637.9 & 41.4 & 17.2 & \textcolor{blue}{\textbf{9.7}} & 111.6 & 2.8 & 0.5 & \textcolor{blue}{\textbf{0.1}} & 8.3 \\
58668.8 & 66.4 & 20.1 & \textcolor{blue}{\textbf{10.1}} & 35.1 & 3.2 & 0.5 & \textcolor{blue}{\textbf{0.1}} & 1.4 \\
58699.7 & \textcolor{blue}{\textbf{8.7}} & 18.2 & 11.7 & 17.9 & \textcolor{blue}{\textbf{0.4}} & 0.4 & 0.2 & 0.3 \\
58790.4 & 18.8 & 15.2 & \textcolor{blue}{\textbf{14.5}} & 14.8 & 1.2 & 0.3 & \textcolor{blue}{\textbf{0.6}} & 0.2 \\
58826.3 & 420.0 & \textcolor{blue}{\textbf{36.9}} & 104.8 & 96.2 & 9.2 & \textcolor{blue}{\textbf{0.6}} & 2.2 & 2.0 \\
58866.2 & \textcolor{blue}{\textbf{11.0}} & 20.0 & 15.8 & 24.6 & \textcolor{blue}{\textbf{0.8}} & 0.7 & 0.8 & 1.4 \\
58909.1 & 73.3 & 55.8 & 89.8 & \textcolor{blue}{\textbf{51.6}} & 2.8 & 1.9 & 3.3 & \textcolor{blue}{\textbf{1.8}} \\
58988.8 & 98.3 & \textcolor{blue}{\textbf{24.5}} & 221.6 & 85.4 & 3.2 & \textcolor{blue}{\textbf{0.5}} & 7.4 & 2.7 \\
59029.9 & 128.0 & \textcolor{blue}{\textbf{31.0}} & 473.1 & 235.7 & 3.3 & \textcolor{blue}{\textbf{0.6}} & 12.6 & 6.3 \\
59061.8 & 135.3 & \textcolor{blue}{\textbf{38.6}} & 288.2 & 128.4 & 3.7 & \textcolor{blue}{\textbf{0.9}} & 8.0 & 3.5 \\
59061.8 & 480.4 & \textcolor{blue}{\textbf{104.5}} & 779.7 & 961.6 & 7.6 & \textcolor{blue}{\textbf{1.6}} & 12.3 & 15.4 \\
59093.4 & 344.9 & \textcolor{blue}{\textbf{137.0}} & 699.4 & 641.4 & 6.7 & \textcolor{blue}{\textbf{2.6}} & 13.6 & 12.6 \\
\end{longtable}

\par}

{\centering
\begin{longtable}{rrlllrlll}
\caption{At 1500 MHz (L-band)} \\
\toprule
MJD & $aicc_p$ & $aicc_b$ & $aicc_l$ & $aicc_m$ & $\chi{^2}_{re_{p}}$ & $\chi{^2}_{re_{b}}$ & $\chi{^2}_{re_{l}}$ & $\chi{^2}_{re_{m}}$ \\
\midrule
\endfirsthead
\toprule
MJD & $aicc_p$ & $aicc_b$ & $aicc_l$ & $aicc_m$ & $\chi{^2}_{re_{p}}$ & $\chi{^2}_{re_{b}}$ & $\chi{^2}_{re_{l}}$ & $\chi{^2}_{re_{m}}$ \\
\midrule
\endhead
\midrule
\multicolumn{9}{r}{Continued on next page} \\
\midrule
\endfoot
\bottomrule
\endlastfoot
59714.4 & 870.3 & 73.6 & 54.7 & \textcolor{blue}{\textbf{50.8}} & 18.4 & 1.4 & 1.0 & \textcolor{blue}{\textbf{0.9}} \\
59728.4 & 379.5 & \textcolor{blue}{\textbf{23.4}} & 39.7 & 36.0 & 10.1 & \textcolor{blue}{\textbf{0.4}} & 0.9 & 0.8 \\
59741.4 & 932.5 & 78.8 & \textcolor{blue}{\textbf{28.0}} & 31.8 & 21.6 & 1.7 & \textcolor{blue}{\textbf{0.5}} & 0.5 \\
59756.3 & 812.3 & 81.9 & \textcolor{blue}{\textbf{11.7}} & 15.1 & 18.8 & 1.7 & \textcolor{blue}{\textbf{0.1}} & 0.2 \\
59776.2 & 870.5 & 36.3 & 36.2 & \textcolor{blue}{\textbf{32.9}} & 20.1 & 0.6 & 0.7 & \textcolor{blue}{\textbf{0.6}} \\
59796.2 & 754.7 & 47.4 & \textcolor{blue}{\textbf{29.5}} & 29.5 & 17.4 & 0.9 & \textcolor{blue}{\textbf{0.5}} & 0.5 \\
59805.1 & 525.9 & 31.1 & 36.1 & \textcolor{blue}{\textbf{25.6}} & 14.1 & 0.6 & 0.8 & \textcolor{blue}{\textbf{0.5}} \\
59813.2 & 1510.4 & 105.1 & \textcolor{blue}{\textbf{87.6}} & 90.7 & 32.0 & 2.1 & \textcolor{blue}{\textbf{1.7}} & 1.8 \\
59852.9 & 1254.1 & 179.0 & \textcolor{blue}{\textbf{50.8}} & 68.8 & 24.5 & 3.4 & \textcolor{blue}{\textbf{0.9}} & 1.2 \\
59862.9 & 988.6 & 134.3 & \textcolor{blue}{\textbf{46.6}} & 51.8 & 20.9 & 2.7 & \textcolor{blue}{\textbf{0.8}} & 0.9 \\
59874.8 & 59.6 & 16.8 & \textcolor{blue}{\textbf{11.3}} & 14.7 & 3.6 & 0.4 & \textcolor{blue}{\textbf{0.2}} & 0.3 \\
59882.8 & 1141.3 & 118.3 & \textcolor{blue}{\textbf{64.6}} & 73.0 & 26.4 & 2.6 & \textcolor{blue}{\textbf{1.4}} & 1.5 \\
59902.9 & 966.7 & 81.2 & \textcolor{blue}{\textbf{22.7}} & 25.3 & 21.4 & 1.6 & \textcolor{blue}{\textbf{0.4}} & 0.4 \\
59909.7 & 844.9 & 105.0 & \textcolor{blue}{\textbf{43.6}} & 66.7 & 19.6 & 2.3 & \textcolor{blue}{\textbf{0.9}} & 1.4 \\
60090.4 & 1577.8 & 310.8 & 238.7 & \textcolor{blue}{\textbf{230.2}} & 23.5 & 4.6 & 3.5 & \textcolor{blue}{\textbf{3.4}} \\
60105.3 & 302.4 & \textcolor{blue}{\textbf{37.0}} & 106.7 & 153.3 & 8.5 & \textcolor{blue}{\textbf{0.8}} & 2.9 & 4.2 \\
60120.2 & 470.9 & \textcolor{blue}{\textbf{58.2}} & 112.8 & 161.2 & 15.0 & \textcolor{blue}{\textbf{1.6}} & 3.4 & 5.1 \\
60133.2 & 548.6 & \textcolor{blue}{\textbf{77.0}} & 98.4 & 112.5 & 15.6 & \textcolor{blue}{\textbf{2.0}} & 2.6 & 3.0 \\
60147.2 & 200.2 & \textcolor{blue}{\textbf{40.8}} & 73.8 & 169.5 & 7.2 & \textcolor{blue}{\textbf{1.2}} & 2.5 & 6.2 \\
60159.1 & 312.6 & \textcolor{blue}{\textbf{23.8}} & 116.2 & 154.2 & 11.4 & \textcolor{blue}{\textbf{0.6}} & 4.0 & 5.6 \\
\end{longtable}

\par}

{\centering
\begin{longtable}{rrlllrlll}
\caption{At 2000 MHz (S-band)} \\
\toprule
MJD & $aicc_p$ & $aicc_b$ & $aicc_l$ & $aicc_m$ & $\chi{^2}_{re_{p}}$ & $\chi{^2}_{re_{b}}$ & $\chi{^2}_{re_{l}}$ & $\chi{^2}_{re_{m}}$ \\
\midrule
\endfirsthead
\toprule
MJD & $aicc_p$ & $aicc_b$ & $aicc_l$ & $aicc_m$ & $\chi{^2}_{re_{p}}$ & $\chi{^2}_{re_{b}}$ & $\chi{^2}_{re_{l}}$ & $\chi{^2}_{re_{m}}$ \\
\midrule
\endhead
\midrule
\multicolumn{9}{r}{Continued on next page} \\
\midrule
\endfoot
\bottomrule
\endlastfoot
59702.4 & 1498.9 & 97.6 & \textcolor{blue}{\textbf{46.8}} & 47.3 & 27.2 & 1.6 & \textcolor{blue}{\textbf{0.7}} & 0.7 \\
59761.2 & 225.0 & 32.6 & \textcolor{blue}{\textbf{17.6}} & 18.2 & 7.1 & 0.8 & \textcolor{blue}{\textbf{0.4}} & 0.3 \\
59762.3 & 1043.2 & 61.0 & 49.0 & \textcolor{blue}{\textbf{44.3}} & 19.6 & 1.0 & 0.8 & \textcolor{blue}{\textbf{0.7}} \\
59886.9 & 4610.2 & 347.2 & \textcolor{blue}{\textbf{90.4}} & 133.4 & 56.9 & 4.2 & \textcolor{blue}{\textbf{1.0}} & 1.6 \\
59896.8 & 4163.5 & 285.6 & \textcolor{blue}{\textbf{117.3}} & 120.9 & 54.0 & 3.6 & \textcolor{blue}{\textbf{1.4}} & 1.5 \\
60084.4 & 2586.3 & 630.5 & \textcolor{blue}{\textbf{235.6}} & 400.3 & 35.4 & 8.6 & \textcolor{blue}{\textbf{3.1}} & 5.4 \\
60137.2 & 777.9 & \textcolor{blue}{\textbf{135.7}} & 150.4 & 204.6 & 16.5 & \textcolor{blue}{\textbf{2.8}} & 3.1 & 4.2 \\
60260.9 & 1838.1 & \textcolor{blue}{\textbf{69.4}} & 104.2 & 105.3 & 39.0 & \textcolor{blue}{\textbf{1.3}} & 2.1 & 2.1 \\
\end{longtable}

\par}

{\centering
\begin{longtable}{rrllrrllr}
\caption{At 5400 MHz (C-band)} \\
\toprule
MJD & $aicc_p$ & $aicc_b$ & $aicc_l$ & $aicc_m$ & $\chi{^2}_{re_{p}}$ & $\chi{^2}_{re_{b}}$ & $\chi{^2}_{re_{l}}$ & $\chi{^2}_{re_{m}}$ \\
\midrule
\endfirsthead
\toprule
MJD & $aicc_p$ & $aicc_b$ & $aicc_l$ & $aicc_m$ & $\chi{^2}_{re_{p}}$ & $\chi{^2}_{re_{b}}$ & $\chi{^2}_{re_{l}}$ & $\chi{^2}_{re_{m}}$ \\
\midrule
\endhead
\midrule
\multicolumn{9}{r}{Continued on next page} \\
\midrule
\endfoot
\bottomrule
\endlastfoot
59714.4 & 856.7 & 153.1 & \textcolor{blue}{\textbf{43.5}} & 53.9 & 18.1 & 3.1 & \textcolor{blue}{\textbf{0.8}} & 1.0 \\
59741.4 & 697.1 & 60.0 & \textcolor{blue}{\textbf{44.5}} & 48.4 & 17.8 & 1.3 & \textcolor{blue}{\textbf{1.0}} & 1.0 \\
59741.4 & 145.1 & \textcolor{blue}{\textbf{16.5}} & 16.5 & 19.7 & 6.7 & \textcolor{blue}{\textbf{0.3}} & 0.4 & 0.5 \\
59756.3 & 615.1 & 73.7 & \textcolor{blue}{\textbf{36.7}} & 63.5 & 14.2 & 1.5 & \textcolor{blue}{\textbf{0.7}} & 1.3 \\
59776.3 & 1214.6 & 146.2 & \textcolor{blue}{\textbf{21.2}} & 33.1 & 23.7 & 2.8 & \textcolor{blue}{\textbf{0.3}} & 0.5 \\
59805.1 & 1347.2 & \textcolor{blue}{\textbf{118.5}} & 121.5 & 191.5 & 27.4 & \textcolor{blue}{\textbf{2.3}} & 2.4 & 3.8 \\
59813.2 & 531.0 & 76.9 & \textcolor{blue}{\textbf{74.8}} & 124.0 & 14.2 & 1.9 & \textcolor{blue}{\textbf{1.8}} & 3.2 \\
59852.9 & 808.9 & \textcolor{blue}{\textbf{54.1}} & 158.9 & 193.0 & 20.6 & \textcolor{blue}{\textbf{1.2}} & 3.9 & 4.8 \\
59862.9 & 831.1 & \textcolor{blue}{\textbf{84.4}} & 249.8 & 486.5 & 19.2 & \textcolor{blue}{\textbf{1.8}} & 5.7 & 11.4 \\
59874.8 & 1201.3 & \textcolor{blue}{\textbf{121.8}} & 188.2 & 258.0 & 24.4 & \textcolor{blue}{\textbf{2.4}} & 3.7 & 5.2 \\
59882.8 & 978.0 & 192.4 & \textcolor{blue}{\textbf{177.5}} & 368.3 & 19.9 & 3.8 & \textcolor{blue}{\textbf{3.5}} & 7.5 \\
59902.9 & 515.0 & \textcolor{blue}{\textbf{81.0}} & 139.5 & 271.2 & 13.1 & \textcolor{blue}{\textbf{1.9}} & 3.4 & 6.9 \\
59909.7 & 795.0 & 233.2 & \textcolor{blue}{\textbf{127.8}} & 194.9 & 17.6 & 5.1 & \textcolor{blue}{\textbf{2.7}} & 4.2 \\
60090.4 & 162.5 & \textcolor{blue}{\textbf{27.5}} & 65.4 & 112.2 & 5.1 & \textcolor{blue}{\textbf{0.6}} & 1.9 & 3.4 \\
60105.3 & 65.8 & \textcolor{blue}{\textbf{22.4}} & 23.8 & 62.7 & 3.2 & \textcolor{blue}{\textbf{0.7}} & 0.9 & 2.9 \\
60133.2 & 231.6 & \textcolor{blue}{\textbf{72.4}} & 104.5 & 227.2 & 6.5 & \textcolor{blue}{\textbf{1.9}} & 2.8 & 6.4 \\
60147.2 & 100.1 & 41.0 & \textcolor{blue}{\textbf{31.7}} & 43.1 & 4.2 & 1.4 & \textcolor{blue}{\textbf{1.1}} & 1.5 \\
60159.1 & 260.2 & 67.6 & \textcolor{blue}{\textbf{62.9}} & 192.1 & 8.2 & 1.9 & \textcolor{blue}{\textbf{1.8}} & 6.1 \\
\end{longtable}

\par}


\end{document}